\documentclass[aip,jcp,reprint,amsmath,amssymb]{revtex4-2}
\usepackage{dcolumn, bm}
\usepackage{graphicx, tikz, orcidlink}
\usepackage{epstopdf}
\usepackage{ascmac}
\usepackage[normalem]{ulem}
\usepackage{cancel}
\usepackage{here}
\usepackage{multirow}
\usepackage{comment}

\DeclareMathOperator{\tr}{tr}

\usepackage[]{hyperref}
\usepackage{xcolor}

\begin{document}
\title{Hierarchical equations of motion for multiple baths (HEOM-MB) and their application to Carnot cycle}
\date{Last updated: \today}

\author{Shoki Koyanagi \orcidlink{0000-0002-8607-1699}}\email[Authors to whom correspondence should be addressed: ]{koyanagi.syoki.36z@st.kyoto-u.jp and tanimura.yoshitaka.5w@kyoto-u.jp}\affiliation{Department of Chemistry, Graduate School of Science,
Kyoto University, Kyoto 606-8502, Japan}

\author{Yoshitaka Tanimura \orcidlink{0000-0002-7913-054X}}
\email[Authors to whom correspondence should be addressed: ]{koyanagi.syoki.36z@st.kyoto-u.jp and tanimura.yoshitaka.5w@kyoto-u.jp}
\affiliation{Department of Chemistry, Graduate School of Science,
Kyoto University, Kyoto 606-8502, Japan}

\begin{abstract}
We have developed a computer code for the thermodynamic hierarchical equations of motion derived from a spin subsystem coupled to multiple Drude baths at different temperatures, which are connected to or disconnected from the subsystem as a function of time.  The code can simulate the reduced dynamics of the subsystem under isothermal, isentropic, thermostatic, and entropic conditions. The extensive and intensive thermodynamic variables are calculated as physical observables,  and Gibbs and Helmholtz energies are evaluated as intensive and extensive work. The energy contribution of the system--bath interaction is evaluated separately from the subsystem using the hierarchical elements of the HEOM. The accuracy of the calculated results for the equilibrium distribution and the two-body correlation functions are assessed by contrasting the results with those obtained from the time-convolution-less Redfield equation. It is shown that the Lindblad master equation is inappropriate for thermodynamic description of a spin--boson system.
Non-Markovian effects in thermostatic processes are investigated by sequentially turning on and off the baths at different temperatures with different switching times and system--bath coupling. In addition, the Carnot cycle is simulated under quasi-static conditions.
To analyze the work done for the subsystem in the cycle, thermodynamic work diagrams are plotted as functions of intensive and extensive variables.
The C++ source codes are provided as supplementary material.
\end{abstract}
\maketitle

\section{Introduction}
\label{sec.intro}
In the early days of NMR\cite{Bloch1953} and optical spectroscopy,\cite{PhysRev.159.208,MOLLOW1969464} the Markov assumption, in which the correlation time of the noise arising from the environment is assumed to be ultra-short, was introduced to account for the linewidths of continuous wave (CW) spectra due to relaxation. 
Since then, the Markov assumption has  been a popular assumption employed for system--bath (SB) models, for example, to derive the reduced equations of motion, such as 
the Redfield equation.\cite{REDFIELD19651} 

In the derivation of these equations, the positivity condition,\cite{davies1976quantum,spohn1980kinetic,pechukas1994reduced} where all diagonal elements of the subsystem density matrix must be positive, is violated without further approximations such as the rotating wave approximation (RWA), which ignores the non-resonant components of the relaxation operator, and the factorization assumption (FA), in which the bath state remains in its thermal equilibrium state as described by its Boltzmann distribution. In regions where the thermal energy is comparable to the excitation energy of the system, the RWA leads to a poor description of the quantum entanglement between the subsystem and the bath (bathentanglement).\cite{T20JCP}  It also alters the dynamics of the subsystem in an inappropriate manner.\cite{DT10PRL,T15JCP,KT24JCP3}  

It is the Markovian assumption that undermines the description of open quantum dynamics, and the RWA and FA were introduced only to restore the positivity problem created by the mathematical but unphysical Markovian assumption. As is clear from the description of the Feynman--Vernon influence functional,\cite{feynman1963dynamical,T14JCP} the effect of a heat bath consists of fluctuation and dissipation, related by the fluctuation--dissipation theorem.  While the dissipation can be Markovian, under the assumption of an Ohmic spectral distribution function (SDF), the fluctuations must be non-Markovian owing to the constraints of the thermal uncertainty principle described as the Matsubara frequency.\cite{T15JCP,T06JPSJ}  

The easiest way to avoid this problem is to abandon the mathematical Markovian assumption.  In fact, the time-convolution-less (TCL) Redfield equation (or Shibata equation)\cite{shibata1977generalized,chaturvedi1979time} for non-Markovian bath approximates the dynamics reasonably well in the perturbed case,  even without RWA.\cite{T15JCP}

Because many important chemical, physical, and biological phenomena arise from interactions with the environment in non-Markovian and non-perturbative regimes, the hierarchical equations of motion (HEOM)\cite{T20JCP} were derived as a generalization of the stochastic Liouville equation\cite{TK89JPSJ1,T06JPSJ} and the master equation.\cite{T90PRA,IT05JPSJ}  The HEOM (or ``doctor equations''\cite{T21JPCB}) are the reduced equations of motion for a subsystem interacting with a harmonic bath. In principle, the HEOM provide an asymptotic approach to the calculations of various physical quantities with any desired accuracy by adjusting the number of hierarchical elements in what is referred to as a numerically ``exact'' approach,\cite{T20JCP} which can be verified by non-Markovian tests based on exact analytical solutions of the Brownian oscillator.\cite{T15JCP}

Unlike the Redfield equation approach, the HEOM do not require the subsystem to be described in terms of the energy eigenstate to construct the damping operator; it can also be solved in phase space.\cite{T15JCP} Thus, even when the eigenstates are time-dependent owing to the presence of a time-dependent external field, the HEOM can accurately describe the dynamics of the subsystem, including the quantum entanglement described by a hierarchical element. This feature is particularly important for the computation of nonlinear response functions, such as multidimensional spectra.\cite{T06JPSJ,T20JCP} The advantages of the HEOM in thermodynamic exploration have been demonstrated.\cite{KT13JPCB,KT15JCP,KT16JCP,ST20JCP,ST21JPSJ} However, while these results were obtained from isothermal processes, we have recently discovered that a complete description of thermodynamics must include thermostatic processes in addition to isothermal, isentropic, and entropic processes.\cite{KT22JCP1,KT22JCP2}

Thus, we  developed a thermodynamic SB model that can take into account a thermostatic process considering multiple heat baths at different temperatures, allowing their coupling on and off in time.  Then, we derived the thermodynamic quantum Fokker--Planck equation (T-QFPE) in the Wigner  representation\cite{KT24JCP1,KT24JCP2,KT24JCP3} on the basis of the low-temperature quantum Fokker--Planck equations for Ohmic SDF without cutoff.\cite{IT19JCTC}  In these investigations, non-Markovian effects arising from quantum fluctuations described as Matsubara frequencies were considered to identify pure quantum effects in comparison with classical results. In this paper, we consider non-Markovian effects of both fluctuation and dissipation arising from Drude cutoff of the Ohmic SDF.

In Sec.~\ref{sec:T-HEOM}, after explaining the thermodynamic SB Hamiltonian, we present the HEOM for multiple baths (HEOM-MB).
In Sec.~\ref{sec:NumericalExamine} we examine the application of the HEOM-MB code. In Sec.~\ref{eq:thermostatic}, we investigate non-Markovian effects in thermostatic processes by sequentially turning on and off baths at different temperatures with different switching times and system--bath coupling. In Sec.~\ref{sec:Carnot}, we demonstrate the capability of our codes by simulating the quasi-static Carnot cycle by present the work diagram as  functions of intensive and extensive variables. Our concluding remarks are presented in 
Sec.~\ref{sec:conclude}. The C++ source codes for the HEOM-MB are provided in the supplementary material.

\section{The HEOM-MB and thermodynamic variables}
\label{sec:T-HEOM}
\subsection{Hamiltonian for multiple baths model}
\label{sec:modelHamiltonian}

Since the extension to multilevel systems is straightforward, we restrict our modeling here to the simplest spin--boson-based system.\cite{SpinBosonLeggett}
We consider a subsystem A coupled to $N_{\rm B} + 1$ heat baths at different inverse temperatures $\beta_k = 1/k_B T_k$, where $k_B$ is the Boltzmann constant. The total Hamiltonian is expressed as
\begin{align}
\hat{H}_{\rm tot}(t) =\hat{H}_{\rm A}(t) + \sum_{k = 0}^{N_{\rm B}} \left( \hat{H}_{\rm I}^{k} (t) + \hat{H}_{\rm B}^{k} \right),
\label{eq:Htotal}
\end{align}
where $\hat{H}_{\rm A}(t)$, $\hat{H}_{\rm I}^{k} (t)$, and $\hat{H}_{\rm B}^{k}$ are the Hamiltonians of the system, $k$th SB interaction, and $k$th bath, respectively.
For the two-bath case ($N_{\rm B} = 1$), the above model has been used to explore heat flow,\cite{KT15JCP}  a quantum heat engine driven by heat flow,\cite{KT16JCP} and the Carnot engine.\cite{KT22JCP2}

We consider a two-level system (TLS) defined as
\begin{equation}
\label{eq:SystemHamiltonian}
\hat{H}_{\rm A} ( t ) = \hat{H}_{\rm A}^0 - B( t ) \hat{\sigma}_z,
\end{equation}
where $B(t)$ is the isothermal driving field (IDF), $\hat{H}_{\rm A}^0$ is the time-independent part of the subsystem Hamiltonian, and $\hat{\sigma}_{\alpha}$ ($\alpha = x$, $y$, or $z$) are the Pauli matrices. Here, we set $\hat{H}_{\rm A}^0 = E_0 \hat{\sigma}_x$, where $E_0$ is the off-diagonal coupling constant. In the case of a spin system, $B(t)$ corresponds to the longitudinal magnetic field and $E_0$ to  the transverse electric (Stark) field. The Hamiltonians representing the $k$th SB interaction and the $k$th bath are given by\cite{KT22JCP1}
\begin{align}
\hat{H}_{\rm I}^{k}(t) = A_k ( t ) \hat{V}_k \sum_j c_{j}^k \left[ \hat{b}_{j}^k + (\hat{b}_{j}^k)^\dagger \right]
\end{align}
and
\begin{align}
\hat{H}_{\rm B}^{k} = \sum_j \hbar \omega_{j}^k \left[ (\hat{b}_{j}^k)^\dagger \hat{b}_{j}^k + \frac{1}{2} \right],
\end{align}
respectively, where $\hat{V}_k$ is the system operator that describes the coupling to the $k$th bath and $A_k ( t )$ is the adiabatic transition field (ATF), introduced to describe the operation of an adiabatic wall between the system and the $k$th heat bath (e.g., the insertion or removal of the adiabatic wall or attaching or detaching the quantum system to/from the bath). Here, $\omega_{j}^k$, $c_{j}^k$, $\hat{b}_{j}^k$, and $(\hat{b}_{j}^k )^\dagger$ are the frequency, coupling strength, and annihilation and creation operators, 
 respectively, of the $j$th mode of the $k$th bath.

The bath effects on the system are determined by the bath correlation function
$C_k^{\rm (B)} (t) \equiv \langle \hat{X}_k(t) \hat{X}_k(0) \rangle_\mathrm{B}$, where $\hat{X}_k \equiv \sum_j c_{j}^k [ \hat{b}_j^k + ( \hat{b}_j^k )^\dagger ]$ is the collective coordinate of the $k$th bath, and $\langle \ldots \rangle_{\mathrm{B}_k}$ represents the average taken with respect to the canonical density operator of the $k$th bath.
The $k$th bath correlation function is expressed in terms of the $k$th bath spectral distribution function (SDF) $J_k (\omega)$ as
\begin{align}
C_k^{\rm (B)} (t)
= \int_0^\infty d\omega \, J_k(\omega)
\left[ \coth\left( \frac{\beta_k \hbar\omega}{2} \right) \cos(\omega t)
- i \sin(\omega t) \right],
\label{eq:BCF}
\end{align}
where $J_k(\omega) \equiv \sum_{j} (c_{j}^k)^2 \delta ( \omega - \omega_{j}^k )$.
We assume the Drude SDF, expressed as
\begin{eqnarray}
\label{eq:SDF}
J_k ( \omega ) = \frac{\hbar}{\pi} \frac{\gamma_k^2 \omega}{\omega^2 + \gamma_k^2} ,
\end{eqnarray}  
where $\gamma_k$ is the inverse of the noise correlation time of the $k$th bath.

The bath correlation functions are then evaluated as
\begin{eqnarray}
\label{eq:sym-approx0}
\begin{split}
C_k^{\rm (B)} ( t ) & \simeq \frac{\gamma_k}{\beta_k} \left( 1 
+ \sum_{j = 1}^{K_k} \frac{2 \eta_j \gamma_k^2}{\gamma_k^2 - [ \nu_j^k ]^2} \right) e^{- \gamma_k | t |}  \\
&
\quad - \sum_{j = 1}^{K_k} \frac{\gamma_k^2}{\beta_k} \frac{2 \eta_j \nu_j^k}{\gamma_k^2 - [ \nu_j^k ]^2} e^{- \nu_j^k | t |} ,
\end{split}
\end{eqnarray}
where $\nu_j, \eta_j$ and $K_k$ are the $j$th Pad{\'e} frequency, the $j$th Pad{\'e} coefficient, and the number of  Pad{\'e} frequencies for the $k$th bath, respectively. \cite{hu2010communication,zhu2011hierarchical,YanPade10B,YanPade11}  

To simplify the discussion, we consider the case of interaction with only one bath at a time, although it is possible to operate multiple baths simultaneously. Such a treatment is important when studying heat flow.\cite{KT15JCP,KT16JCP}
Thus, the SB coupling strength and temperature of the bath attached to the subsystem are expressed as\cite{KT22JCP1,KT22JCP2}
\begin{eqnarray}
A ( t ) = \sum_{k = 0}^{N_{\rm B}} A_k ( t ),
\end{eqnarray}
and
\begin{eqnarray}
\label{eq:Bathb}
\beta ( t ) = \sum_{k = 0}^{N_{\rm B}} {\beta_k \xi_k ( t )},
\end{eqnarray}
respectively. The temperature of the bath is defined as $T ( t ) = 1 / k_{B} \beta ( t )  $.  

The window function (thermostatic field) is defined as
\begin{eqnarray}
\label{eq:xi}
 \xi_k (t)= \theta(t-t_k)\theta(t_k +\Delta t -t),
\end{eqnarray}
where  $\theta (t)$ is the step function and the time $t_k$ is defined as $t_k = t_0 + k \Delta t$, with  initial time $t_0$ and  duration $\Delta t$.  

\subsection{HEOM-MB}
In the HEOM formalism, the set of equations of motion consists of the auxiliary density operators (ADOs).\cite{T06JPSJ,T20JCP}
As we show in Eq.~\eqref{eq:sym-approx0}, the bath correlation function is written as a linear combination of exponential functions.
To adapt the HEOM formalism, we assume that all $K^k, \gamma_k$, and $\hat{V}_k$ are independent of $k$, and are thus written as $K, \gamma$, and $\hat{V}$, respectively. 

To treat multiple baths, we must consider ADOs for each bath, because the bathentanglement does not disappear quickly, owing to the non-Markovian nature of bath noise. However, assuming that the detached bath never reattaches to the subsystem, the bathentanglement after detachment does not affect the system dynamics. Thus, instead of explicitly dealing with the $(K + 1) \times (N_{\rm B} + 1)$-dimensional hierarchy, we only need one set of hierarchies and reuse the same hierarchies to describe the successive baths by resetting $\hat{\rho}_{\vec{n}} = 0\; (\vec{n} \neq \vec{0})$ before attaching the next bath.

Then, the ADOs introduced in the HEOM-MB are defined by $\hat{\rho}_{\vec{n}} ( t )$, with a set of indices $\vec{n} = ( n_0 , n_1 , \cdots , n_K )$, where $n_{l}$ represents an integer value of zero or greater. The zeroth ADO, $\hat{\rho}_{\vec{0}}(t)$
with $\vec{0} = ( 0, 0, \ldots, 0)$, corresponds to the actual reduced density operator. The HEOM for the IDF and the ATF are then expressed as\cite{KT22JCP1,KT22JCP2}
\begin{align}
\frac{\partial}{\partial t} \hat{\rho}_{\vec{n}} ( t ) &= \left( - \frac{i}{\hbar} \hat{H}_{\rm A}^\times ( t ) - 
\sum_{l=0}^{K} n_l \nu_l (t) \right) \hat{\rho}_{\vec{n}} ( t ) \nonumber \\
& - \frac{i}{\hbar} {A( t )} \sum_{l = 0}^{K} n_{l} \hat{\Theta}_{l}(t) \hat{\rho}_{\vec{n} - \vec{e}_l} ( t ) \nonumber \\
& - \frac{i}{\hbar}  {A( t )} \sum_{l = 0}^{K} \hat{V}^\times \hat{\rho}_{\vec{n} + \vec{e}_l} ( t ),
\label{ModelHEOM}
\end{align}
where $\vec{e}_{l}$ is the ($K$+1)-dimensional unit vector. We introduce a set of fluctuation--dissipation operators as
\begin{equation}
\label{eq:Theta0}
\hat{\Theta}_{0} (t) \equiv \left( \frac{ \gamma}{\beta(t)} + \sum_{m = 1}^{K} \frac{ \eta_m \gamma^2}{\beta(t)} \frac{ 2 \gamma}{\gamma^2 - [ \nu_m (t) ]^2} \right) \hat{V}^\times - \frac{i \hbar \gamma^2}{2} \hat{V}^\circ
\end{equation}
and
\begin{align}
\label{eq:Thetal}
\hat{\Theta}_{l}(t) \equiv - \frac{ \eta_l \gamma^2}{\beta(t)} \frac{2 \nu_l (t) }{\gamma^2 - [ \nu_l (t) ]^2} \hat{V}^\times ,
\end{align}
where $\hat{\mathcal{O}}^\times \hat{\mathcal{P}} = [ \hat{\mathcal{O}} , \hat{\mathcal{P}} ]$ and $\hat{\mathcal{O}}^\circ \hat{\mathcal{P}} = \{ \hat{\mathcal{O}} , \hat{\mathcal{P}} \}$ for arbitrary operators $\hat{\mathcal{O}}$ and $\hat{\mathcal{P}}$, and $[ \hat{\mathcal{O}} , \hat{\mathcal{P}} ]$ and $\{ \hat{\mathcal{O}} , \hat{\mathcal{P}} \}$ are the commutator and the anticommutator.

As the temporal initial conditions, we consider the factorized initial case
$$
\hat{\rho}_{\rm tot}(0) = \hat{\rho}_{\rm A}(0) \prod_{k = 0}^{N_{\rm B}} \frac{e^{- \beta_k \hat{H}_{\rm B}^{k} } }{ {\rm tr}_{\rm B}\{ e^{ - \beta_k \hat{H}_{\rm B}^{k} }\} } ,
$$
where $\hat{\rho}_{\rm A}(t)=\hat{\rho}_{\vec{0}}(t)$ is the reduced density operator of the subsystem. 

Using the zeroth member of the hierarchy, $\hat{\rho}_{\vec{0}} ( t ) $, we define the enthalpy and the internal energy of the subsystem as follows:\cite{KT24JCP1}
\begin{eqnarray}
\label{eq:DefHA}
H_{\rm A} ( t ) = \tr_{\rm A} \left\{ \hat{H}_{\rm A} ( t ) \hat{\rho}_{\vec{0}} ( t ) \right\}
\end{eqnarray}
and
\begin{align}
\label{eq:DefUA}
U_{\rm A} (t) = \tr_{\rm A} \left\{ \hat{H}_{\rm A}^0 ( t ) \hat{\rho}_{\vec{0}} ( t ) \right\}.
\end{align}
From Eq.~\eqref{eq:SystemHamiltonian}, the enthalpy and the internal energy satisfy the time-dependent Legendre transformation expressed as
\begin{eqnarray}
\label{eq:LegendreU-H}
H_{\rm A} ( t ) = U_{\rm A} ( t ) - B ( t ) M_{\rm A} ( t ) ,
\end{eqnarray}
where $M_{\rm A} ( t )$ is magnetization defined as
\begin{eqnarray}
\label{eq:M}
M_{\rm A} ( t ) = \tr \{ \hat{\sigma}_z \hat{\rho}_{\vec{0}} ( t ) \} .
\end{eqnarray}
Because the higher-order hierarchical elements store the information of the SB coupling, we can also evaluate the SB interaction energy as follows:\cite{KT16JCP,ST20JCP,ST21JPSJ,KT22JCP1,KT22JCP2}
\begin{eqnarray}
\label{eq:InteractionEnergy}
H_{{\rm I}} (t) &&\equiv
\tr_{\rm tot} \left\{ \sum_{k = 0}^{N_{\rm B}} \hat{H}_{\rm I}^k ( t ) \hat{\rho}_{\rm tot} ( t ) \right\} \nonumber \\
&&= A ( t ) \sum_{l = 0}^K \tr \left\{ \hat{V} \hat{\rho}_{\vec{e}_l} ( t ) \right\} .
\end{eqnarray}
The expectation value of the bath energy is defined as $H_{\rm B} (t)\equiv {\rm{tr}}\{   { \hat H}_{\rm B} {\hat \rho}_{\rm tot}(t)\}$.  From $\partial H_{\rm B} (t) / \partial t = i \int_{t_0}^{t} {\rm{tr}}\{ [\hat H_{\rm I}, \hat H_{\rm B}] \hat \rho_{\rm tot} ( t ) \}\, dt/\hbar$, which is obtained from $\partial \hat{\rho}_{\rm tot} ( t ) / \partial t = [ \hat{H}_{\rm tot} ( t ) , \hat{\rho}_{\rm tot} ( t ) ] / i \hbar$, we have (see Appendix~\ref{sec:BathEnergy})
\begin{align}
\label{eq:expectEB}
\frac{\partial}{\partial t}
H_{\rm B}^{\rm neq} ( t ) &= A ( t ) \sum_{l = 0}^{K} \nu_{l} ( t )
\tr_{\rm A} \{ \hat{V} \hat{\rho}_{\vec{e}_l} ( t ) \} \nonumber \\
& + A^2 ( t ) \gamma^2 \tr_{\rm A} \left\{ \hat V^2 \hat{\rho}_{\vec{0}} ( t ) \right\}.
\end{align}
The change in bath energy $\Delta H_{\rm B} (t)$ can be evaluated from the numerical integration of the above.\cite{KT16JCP,ST20JCP,ST21JPSJ}  Thus, we have $\Delta H_{\rm tot} (t) = \Delta H_{\rm A} (t) + \Delta H_{\rm I} (t)+ \Delta H_{\rm B} (t)$, where $\Delta H_{\alpha} (t) \equiv H_{\alpha} (t) - H_{\alpha} (t_0)$ for $\alpha=A, B, I$, and tot.

\subsection{Intensive and extensive work: thermodynamic potentials}
We have previously considered thermodynamic Brownian models with Ohmic SDFs and derived the laws of thermodynamics for entropic potentials, based on the dimensionless minimum work principle for a subsystem.\cite{KT24JCP1,KT24JCP2,KT24JCP3} In that study, multiple baths with small temperature differences were assumed to be switched on and off more slowly than the characteristic time scale of the subsystem, and the bath change process was approximated by considering the temperature as a function of time. 
In non-Brownian and non-Ohmic systems, however, such an approximation is generally not valid, because the work done by an external field includes the contribution of the subsystem part of the SB interaction, which depends on the noise correlation time ($\approx 1/\gamma$ for the Drude case).\cite{ST20JCP}

In the present spin--boson case, we therefore treat the multiple-bath model explicitly by attaching and detaching the baths in a stepwise fashion and consider the work not from the subsystem but from the total system, defined as
\begin{eqnarray}
\frac{d W^{int}_{\rm tot} ( t )}{d t}
= \tr_{\rm tot} \left\{ \frac{\partial \hat{H}_{\rm tot} ( t )}{\partial t} \hat{\rho}_{\rm tot} ( t ) \right\}.
\end{eqnarray}
where $W^{int}_{\rm tot} ( t )$ is the intensive work. While $B ( t )$ is the intensive variable for the subsystem, $A ( t )$ is that for the total system. The conjugated extensive variable is defined using the ADOs as
\begin{eqnarray}
\label{eq:D}
D_{\rm I} ( t ) = \sum_{j = 0}^K \tr \{ \hat{\sigma}_x \hat{\rho}_{\vec{e}_j} ( t ) \},
\end{eqnarray}
which we call strain.\cite{KT22JCP1,KT22JCP2} 
The time derivative of the total intensive work is then expressed as
\begin{eqnarray}
\label{eq:DefWint-tot}
\frac{d W_{\rm tot}^{int} ( t )}{d t}
= - M_{\rm A}  ( t ) \frac{d B ( t )}{d t} - D_{\rm I} ( t ) \frac{d A ( t )}{d t} .
\end{eqnarray}

Although most thermodynamic cycles do not include a thermostatic process, we have found that a complete thermodynamic description requires a thermostatic process. Thus, we consider the total dimensionless (DL) intensive work expressed as\cite{KT24JCP1, KT24JCP2}
\begin{eqnarray}
\label{eq:DefDLWint}
\frac{d \tilde{W} _{\rm tot} ^{int} ( t )}{d t}
= \tr_{\rm tot} \left\{ \frac{\partial \hat{\tilde{H}}_{\rm tot} ( t )}{d t} \hat{\rho}_{\rm tot} ( t ) \right\} ,
\end{eqnarray}
where the total DL Hamiltonian is defined as
\begin{eqnarray}
\hat{\tilde{H}}_{\rm tot} ( t ) = \beta ( t ) \hat{H}_{\rm A} ( t )
+ \sum_{k = 0}^{N_{\rm B}} \beta_k \left( \hat{H}_{\rm I}^{k} (t) + \hat{H}_{\rm B}^{k} \right) .
\end{eqnarray}
In terms of the intensive and extensive variables, Eq.~\eqref{eq:DefDLWint} is expressed as
\begin{align}
\frac{d \tilde{W}_{\rm tot}^{int} ( t )}{d t}
= U_{\rm A} ( t ) \frac{d \beta ( t )}{d t} - M _{\rm A}  ( t ) \frac{d \tilde{B} ( t )}{d t} 
- \tilde{D}_{\rm I} ( t ) \frac{d A ( t )}{d t} ,
\end{align}
where $\tilde{B} ( t ) \equiv \beta ( t ) B ( t )$ and $\tilde{D}_{\rm I} ( t ) \equiv \beta ( t ) D_{\rm I} ( t )$.

In evaluating the thermodynamic potential, it is necessary to distinguish whether the work is done by an intensive variable or an extensive variable.  Thus, we introduce the intensive and extensive work defined as
\begin{align}
\label{eq:DefWext}
\frac{d W_{\rm tot}^{ext} ( t )}{d t}  = \frac{d W_{\rm tot}^{int} ( t )}{d t} 
+ \frac{d}{dt}[ B(t) M ( t ) ]  + \frac{d}{dt}[ A(t) D_{\rm I} ( t ) ],  
\end{align}
which is the time-dependent Legendre transformation between $W_{\rm tot}^{int} ( t )$ and $W_{\rm tot}^{ext} ( t )$.\cite{KT24JCP1,KT24JCP2} 
Accordingly we introduce the DL extensive work defined as
\begin{align}
\label{eq:DefDLWext}
\frac{d \tilde{W}_{\rm tot}^{ext} ( t )}{d t}
= \frac{d \tilde{W}_{\rm tot}^{int} ( t )}{d t} + \frac{d}{d t} [ \tilde{B} ( t ) M ( t ) ]
+ \frac{d}{d t} [ A ( t ) \tilde{D}_{\rm I} ( t ) ] . 
\end{align}
The first law of thermodynamics is now expressed as
\begin{eqnarray}
\label{eq:DefDLQext}
\frac{d \tilde{Q}_{\rm tot}^{ext} ( t )}{d t}
= \frac{d \tilde{U}_{\rm A} ( t )}{d t} - \frac{d \tilde{W}_{\rm tot}^{ext} ( t )}{d t} ,
\end{eqnarray}
where $\tilde{U}_{\rm A} ( t ) \equiv \beta ( t ) U_{\rm A} ( t )$ is the DL internal energy.

\subsection{Massieu--Planck potentials}
\label{Massieu}
The DL Massieu and Planck potentials are defined as the minimum values of the quasi-static dimensionless (or entropic) work and heat expressed as\cite{Guggenheim1986,massieu1869,Planck1922,FreeEntropyPlanes2002}
\begin{eqnarray}
\label{eq:MinimumPrinciple0}
\tilde{W}_{\rm tot}^{int} \geq - \Delta \Xi_{\rm tot}^{\rm qst} 
\end{eqnarray}
and
\begin{eqnarray}
\label{eq:MinimumExtWork0}
\tilde{W}_{\rm tot}^{ext} \geq -\Delta \Phi_{\rm tot}^{\rm qst},
\end{eqnarray}
respectively. The latter inequality is a generalization of the Kelvin--Planck statement, often used as a definition of the second law of thermodynamics, for isothermal processes. From Eq.~\eqref{eq:DefDLQext}, the  Clausius entropy (C-entropy)  is defined as
\begin{eqnarray}
\label{eq:MaximumDLHeat0}
\tilde{Q}_{\rm tot}^{ext}  \leq \Delta \Gamma_{\rm tot}^{\rm qst},
\end{eqnarray}
which we call the ``principle of maximum dimensionless heat generation'' and which is equivalent to  the ``principle of maximum C-entropy.''

The equalities in Eqs.~\eqref{eq:MinimumPrinciple0} and~\eqref{eq:MinimumExtWork0} hold when the external perturbation of the entire system is quasi-static. Therefore, the entropic potentials can be evaluated as a quasi-static intensive and extensive work as $\Delta \Xi_{\rm tot}^{\rm qst}(t)=-[\tilde{W}_{\rm tot}^{int} (t)]^{\rm qst}$ and $\Delta \Phi_{\rm tot}^{\rm qst}(t)=-[\tilde{W}_{\rm tot}^{ext} (t)]^{\rm qst}$. The C-entropy can be evaluated accordingly.

\begin{table*}[!t]
\caption{\label{table:DLPotential} Total differential expressions for the quasi-static (qst.) entropic potentials as functions of the intensive variables $\beta^{\rm qst}(t)$, $B^{\rm qst} ( t )$,  and $A^{\rm qst} ( t )$ and the extensive variables $U_{\rm A}^{\rm qst}(t)$, $\tilde{M}_{\rm A}^{\rm qst} ( t )$, and $\tilde{D}^{\rm qst} ( t )$ for fixed $\tilde{\gamma}^{\rm qst}(t) \equiv \beta(t) \gamma^{\rm qst} ( t )$. Entropy has two definitions, depending on whether the work variable is intensive or extensive.
Of these dimensionless entropies, the most commonly used one, which we call Clausius entropy (C-entropy) and which involves only extensive variables, is expressed as $\Gamma_{\rm A}^{\rm qst} [U_{\rm A}^{\rm qst}, M_{\rm A}^{\rm qst} , \tilde{D}^{\rm qst} ]$, 
while the less common one, which we call Boltzmann entropy (B-entropy),  is expressed as $\Lambda_{\rm A}^{\rm qst} [ U_{\rm A}^{\rm qst}, \tilde{B}^{\rm qst} , A^{\rm qst} ]$. The potentials are related by the Legendre transformations as shown.}
\begin{ruledtabular}
\begin{tabular}{llcc}
Qst. Potential & Differential form & Natural var. & Legendre transformation 
\\
\hline
Massieu & 
$d \Phi_{\rm tot}^{\rm qst}  = - U_{\rm A}^{\rm qst} d \beta^{\rm qst}  - \tilde{B}^{\rm qst} d M_{\rm A}^{\rm qst} - A^{\rm qst} d \tilde{D}_{\rm I}^{\rm qst}$
& $\beta^{\rm qst} , M_{\rm A}^{\rm qst}, \tilde{D}_{\rm I}^{\rm qst}$ & $\cdots$ 
\\
Planck  & 
$d \Xi_{\rm tot}^{\rm qst}  = - U_{\rm A}^{\rm qst} d \beta^{\rm qst} + M_{\rm A}^{\rm qst} d \tilde{B}^{\rm qst}+ \tilde{D}_{\rm I}^{\rm qst} d A^{\rm qst}$
& $\beta^{\rm qst} , \tilde{B}^{\rm qst}, A^{\rm qst}$ 
& $\Xi_{\rm tot}^{\rm qst} = \Phi_{\rm tot}^{\rm qst} + \tilde{B}^{\rm qst} M_{\rm A}^{\rm qst} + A^{\rm qst} \tilde{D}_{\rm I}^{\rm qst}$
\\
C-Entropy 
& $d \Gamma_{\rm tot} ^{\rm qst} = \beta^{\rm qst} d U_{\rm A}^{\rm qst} - \tilde{B}^{\rm qst} d M_{\rm A}^{\rm qst} - A^{\rm qst} d \tilde{D}_{\rm I}^{\rm qst}$
& $U_{\rm A}^{\rm qst} , M_{\rm A}^{\rm qst}, \tilde{D}_{\rm I}^{\rm qst}$
& $\Gamma_{\rm tot} ^{\rm qst} = \Phi_{\rm tot} ^{\rm qst} + \beta^{\rm qst} U^{\rm qst}_{\rm A} $
\\
B-Entropy 
& $d \Lambda_{\rm tot}^{\rm qst}  = \beta^{\rm qst} d U_{\rm A}^{\rm qst} + M_{\rm A}^{\rm qst} d \tilde{B}^{\rm qst} + \tilde{D}_{\rm I}^{\rm qst} d A^{\rm qst}$
& $U_{\rm A}^{\rm qst} , \tilde{B}^{\rm qst}, A^{\rm qst}$
& $\Lambda_{\rm tot}^{\rm qst}  =  \Xi_{\rm tot}^{\rm qst} + \beta^{\rm qst} U^{\rm qst}_{\rm A} $
\end{tabular}
\end{ruledtabular}

\end{table*}

\begin{table*}
\caption{\label{table:Potential} Total differential expressions for the quasi-static (qst.) thermodynamic potentials as functions of the intensive variables $T^{\rm qst}$, $B^{\rm qst}$, and $A^{\rm qst}$ and the extensive variables $S_{\rm tot}^{\rm qst}$, $M_{\rm A}^{\rm qst}$, and $D_{\rm I}^{\rm qst}$ for fixed $\tilde{\gamma}^{\rm qst} ( t )$. We note that while the Helmholtz and Gibbs energies are the total system quantities, the internal energy and the enthalpy are the subsystem quantities.}
\begin{ruledtabular}
\begin{tabular}{llcc}
Qst. Potential & Differential form & Natural var. & Legendre transformation
\\
\hline
Helmholtz 
& $d F_{\rm tot}^{\rm qst} = - S_{\rm tot}^{\rm qst} d T^{\rm qst} + B^{\rm qst} d M_{\rm A}^{\rm qst} + A^{\rm qst} d D_{\rm I}^{\rm qst}$
& $T^{\rm qst} , M_{\rm A}^{\rm qst} , D_{\rm I}^{\rm qst}$
& $\cdots$
\\
Gibbs
& $d G_{\rm tot}^{\rm qst} = - S_{\rm tot}^{\rm qst} d T^{\rm qst} - M_{\rm A}^{\rm qst} d B^{\rm qst}
- D_{\rm I}^{\rm qst} d A^{\rm qst}$
& $T^{\rm qst} , B^{\rm qst} , A^{\rm qst}$
& $G_{\rm tot}^{\rm qst} = F_{\rm tot}^{\rm qst} - B^{\rm qst} M_{\rm A}^{\rm qst}
- A^{\rm qst} D_{\rm I}^{\rm qst}$
\\
Internal Energy
& $d U_{\rm A}^{\rm qst} = T^{\rm qst} d S_{\rm tot}^{\rm qst} + B^{\rm qst} d M_{\rm A}^{\rm qst}
+ A^{\rm qst} d D_{\rm I}^{\rm qst}$
& $S_{\rm tot}^{\rm qst} , M_{\rm A}^{\rm qst} , D_{\rm I}^{\rm qst}$
& $U_{\rm A}^{\rm qst} = F_{\rm tot}^{\rm qst} + T^{\rm qst} S_{\rm tot}^{\rm qst}$
\\
Enthalpy
& $d H_{\rm A}^{\rm qst} = T^{\rm qst} d S_{\rm tot}^{\rm qst} - M_{\rm A}^{\rm qst} d B^{\rm qst}
- D_{\rm I}^{\rm qst} d A^{\rm qst}$
& $S_{\rm tot}^{\rm qst} , B^{\rm qst} , A^{\rm qst}$
& $H_{\rm A}^{\rm qst} = G_{\rm tot}^{\rm qst} + T^{\rm qst} S_{\rm tot}^{\rm qst}$
\\
\end{tabular}
\end{ruledtabular}

\end{table*}

These entropic potentials can be related by a time-dependent Legendre transformation obtained from Eqs.~\eqref{eq:DefDLWext} and~\eqref{eq:DefDLQext} as follows:
\begin{eqnarray}
\label{eq:LegendreP-M}
\Phi_{\rm tot}^{\rm qst} ( t ) = \Xi_{\rm tot}^{\rm qst} ( t ) - \tilde{B}^{\rm qst} ( t ) M_{\rm A}^{\rm qst} ( t ) 
- A^{\rm qst} ( t ) \tilde{D}_{\rm I}^{\rm qst} ( t )
\end{eqnarray}
and
\begin{eqnarray}
\label{eq:LegendreM-E}
\Gamma_{\rm tot}^{\rm qst} ( t ) = \Phi_{\rm tot}^{\rm qst} ( t ) 
+ \beta^{\rm qst} ( t ) U_{\rm A}^{\rm qst} ( t ).
\end{eqnarray}
The  Boltzmann entropy (B-entropy) is evaluated as  
\begin{eqnarray}
\label{eq:LegendreP-S}
\Lambda_{\rm tot}^{\rm qst} ( t ) = \Xi_{\rm tot}^{\rm qst} ( t ) 
+ \beta^{\rm qst} ( t ) U_{\rm A}^{\rm qst} ( t ).
\end{eqnarray}
We summarize the DL entropic potentials as functions of the natural variables in total differential form in Table~\ref{table:DLPotential}.

Because $\beta ( t )$ is assumed to change only when $A ( t ) = 0$, the Legendre transformations [Eqs.~\eqref{eq:LegendreM-E} and~\eqref{eq:LegendreP-S}] hold only under this condition.  
Note that by fixing $\tilde{\gamma}^{\rm qst} ( t ) = \beta^{\rm qst} ( t ) \hbar \gamma^{\rm qst} ( t )$, we can ensure that (see Appendix~\ref{sec:TDEnthalpy})
\begin{eqnarray}
\label{eq:Planck-U}
U_{\rm A}^{\rm qst} ( t ) = - \left. \left( \frac{\partial \Xi_{\rm tot}^{\rm qst} ( t )}{\partial \beta^{\rm qst} ( t )}
\right) \right|_{\tilde{B}^{\rm qst} , A^{\rm qst} , \tilde{\gamma}^{\rm qst}} ,
\end{eqnarray}
even when $A^{\rm qst} ( t ) \neq 0$, where $\gamma ( t ) = \sum_{k = 0}^{N_{\rm B}} \gamma_k \xi_k ( t )$. In this case, the Legendre transformation remain applicable regardless of the value of $A(t)$.

As the difference between the total DL intensive work and Planck potential, we introduce
the entropy production expressed as\cite{KT24JCP2} 
\begin{eqnarray}
\label{eq:loss}
\Sigma_{\rm tot} = \tilde{W}_{\rm tot}^{int} + \Delta \Xi_{\rm tot}^{\rm qst}.
\end{eqnarray}

In isothermal processes, the DL minimum work principle [Eq.~\eqref{eq:MinimumPrinciple0}] is reduced to the Kelvin--Planck statement, expressed as $W_{\rm tot}^{ext} \geq \Delta F_{\rm tot}^{\rm qst}$ and $W_{\rm tot}^{int} \geq \Delta G_{\rm tot}^{\rm qst}$, where
\begin{eqnarray}
\label{eq:Planck-Gibbs}
G_{\rm tot}^{\rm qst} ( t ) \equiv - \frac{\Xi_{\rm tot}^{\rm qst} ( t )}{\beta^{\rm qst} ( t )} 
\end{eqnarray}
and
\begin{eqnarray}
\label{eq:Massieu-Helmholtz}
F_{\rm tot}^{\rm qst} ( t ) \equiv - \frac{\Phi_{\rm tot}^{\rm qst} ( t )}{\beta^{\rm qst} ( t )} 
\end{eqnarray}
are the quasi-static Gibbs and Helmholtz energies, respectively. 

We can calculate the quasi-static entropy from the Gibbs energy as
\begin{eqnarray}
S_{\rm tot}^{\rm qst} ( t ) = -
\left( \frac{\partial G_{\rm tot}^{\rm qst} ( t )}{\partial T^{\rm qst} ( t )} \right)_{B^{\rm qst} , A^{\rm qst} , \tilde{\gamma}^{\rm qst}} ,
\end{eqnarray}
which satisfies $S_{\rm tot}^{\rm qst} ( t ) = k_{B} \Gamma_{\rm tot}^{\rm qst} ( t )$.

We summarize the DL thermodynamic potentials as functions of the natural variables in total differential form in Table~\ref{table:Potential}.

\section{Numerical tests for isothermal spin--boson system}
\label{sec:NumericalExamine}

When the spin is placed in a dissipative environment, not only the populations but also the coherence of the spin states become entangled with the bath, and their dynamical behavior is altered depending on the temperature, noise correlation time, and coupling strength of the bath.\cite{T14JCP,T20JCP} Therefore, the study of simple population relaxation dynamics alone does not validate the description of open quantum dynamics.  Moreover, since the true equilibrium state of the subsystem is entangled with the bath and is not the Boltzmann distribution of the subsystem, and since it takes a long time for the effects of relaxation to become apparent, the difference in relaxation mechanisms is not clear from looking only at the initial decay of the population relaxation process.

Non-Markovian tests for a harmonic Brownian system have been developed to clarify these points.\cite{T15JCP}  
Using the energy-eigenstate representation of the harmonic oscillator, non-Markovian tests have also been performed for the HEOM for the discretized energy states, even in the Ohmic case.\cite{KT24JCP3}  Although there are no analytical solutions for the spin--boson system, here we perform the simulations following the protocol of the  non-Markovian tests for the descriptions of  (i) the equilibrium distribution, (ii) the linear response function $R^{(1)} ( t ) = (i/\hbar)\tr_{\rm tot} \{ [ \hat{\sigma}_x ( t ) , \hat{\sigma}_x ] \hat{\rho}_{\rm tot}^{\rm eq} \}$, and (iii) the symmetric correlation function $C ( t ) = \tr_{\rm tot} \{ \{ \hat{\sigma}_x ( t ) , \hat{\sigma}_x \} \hat{\rho}_{\rm tot}^{\rm eq} \} / 2$, 
where $\hat{\sigma}_x ( t )$ is the Heisenberg representation of $\hat{\sigma}_x$.  Note that unlike a harmonic Brownian system, symmetric and antisymmetric correlation functions for the spin--boson system depend on both the coupling strength and the temperature. 

For the numerical integration of the HEOM-MB [Eq.~\eqref{ModelHEOM}] we use the Runge--Kutta--Fehlberg method, and the time step size $\delta t$ is automatically determined by the algorithm (see Appendix~\ref{sec:RKF}).
To show that the TCL-Redfield and Lindblad master equations derived under the perturbative SB coupling approximation ignoring the bathentanglement (see Appendix~\ref{sec:Redfield-master}) are inappropriate for thermodynamic description, we also conduct the same tests for these equations.
These equations are integrated by the fourth-order Runge--Kutta method with  time step $\delta t = 0.001$.

For the simulations, we set the external forces to be constant and set $B ( t ) = 1.0$, $E_0 = 0.0$, $A ( t ) = A$, and $\beta ( t ) = \beta$ for fixed $\gamma = 1.0$. The truncation numbers $N$ and $K$ are listed in Table~\ref{table:NK}. To illustrate the anomalous feature of the spin--boson system with the Ohmic SDF, we set $\gamma = 100$ 
for the simulation and compare the results with those obtained from the Lindblad master equation in Appendix~\ref{sec:OhmicLimit}.

\begin{table}
\caption{\label{table:Eq} Expectation values $\langle \hat{\sigma}_z \rangle = \tr_{\rm A} \{ \hat{\sigma}_z \hat{\rho}_{\rm A}^{\rm eq} \}$ in the equilibrium state computed using the HEOM-MB and the TCL-Redfield equation for $A=1$ are shown for  bath temperatures $\beta = 0.1$ (hot), $\beta = 1.0$ (intermediate), and $\beta = 10.0$ (cold).}
\begin{ruledtabular}
\begin{tabular}{cccc}
$\beta$ & \hspace{0.25cm} HEOM-MB \hspace{0.25cm} & \hspace{0.25cm} TCL-Redfield \hspace{0.25cm} 
\\
\hline 
$0.1$ & $9.65 \times 10^{-2}$ & $9.97 \times 10^{-2}$ 
\\
$1.0$ & $0.619$ & $0.762$ 
\\
$10.0$ & $0.891$ & $1.00$ 
\end{tabular}
\end{ruledtabular}
\end{table}
To obtain the equilibrium distributions, we compute the expectation value of $\hat{\sigma}_z$ using the HEOM-MB and the TCL-Redfield equation at equilibrium.  The results are shown in Table~\ref{table:Eq}. We set the SB coupling strength $A = 1.0$ and compute the equilibrium distribution for  three inverse temperatures $\beta = 0.1$ (hot), $\beta = 1.0$ (intermediate), and $\beta = 10.0$ (cold). In the high-temperature case, the results obtained from the HEOM-MB and the TCL-Redfield equation are nearly the same because the SB entanglement effect becomes small in this case, whereas in the low-temperature case, $\langle \hat{\sigma}_z \rangle$  obtained from the HEOM-MB is smaller than that obtained from the TCL-Redfield equation, owing to the bathentanglement. We also note that the values of $\langle \hat{\sigma}_z \rangle$ obtained from the TCL-Redfield results almost agree with the result obtained from the Boltzmann distribution of the subsystem, $\tr_{\rm A} \{ \hat{\sigma}_z e^{-\beta \hat{H}_{\rm A}} \} / \tr_{\rm A} \{ e^{- \beta \hat{H}_{\rm A}} \}$, which indicates that 
the TCL-Redfield equation ignores the effects of bathentanglement.

\begin{figure}
\includegraphics[width=8.2cm]{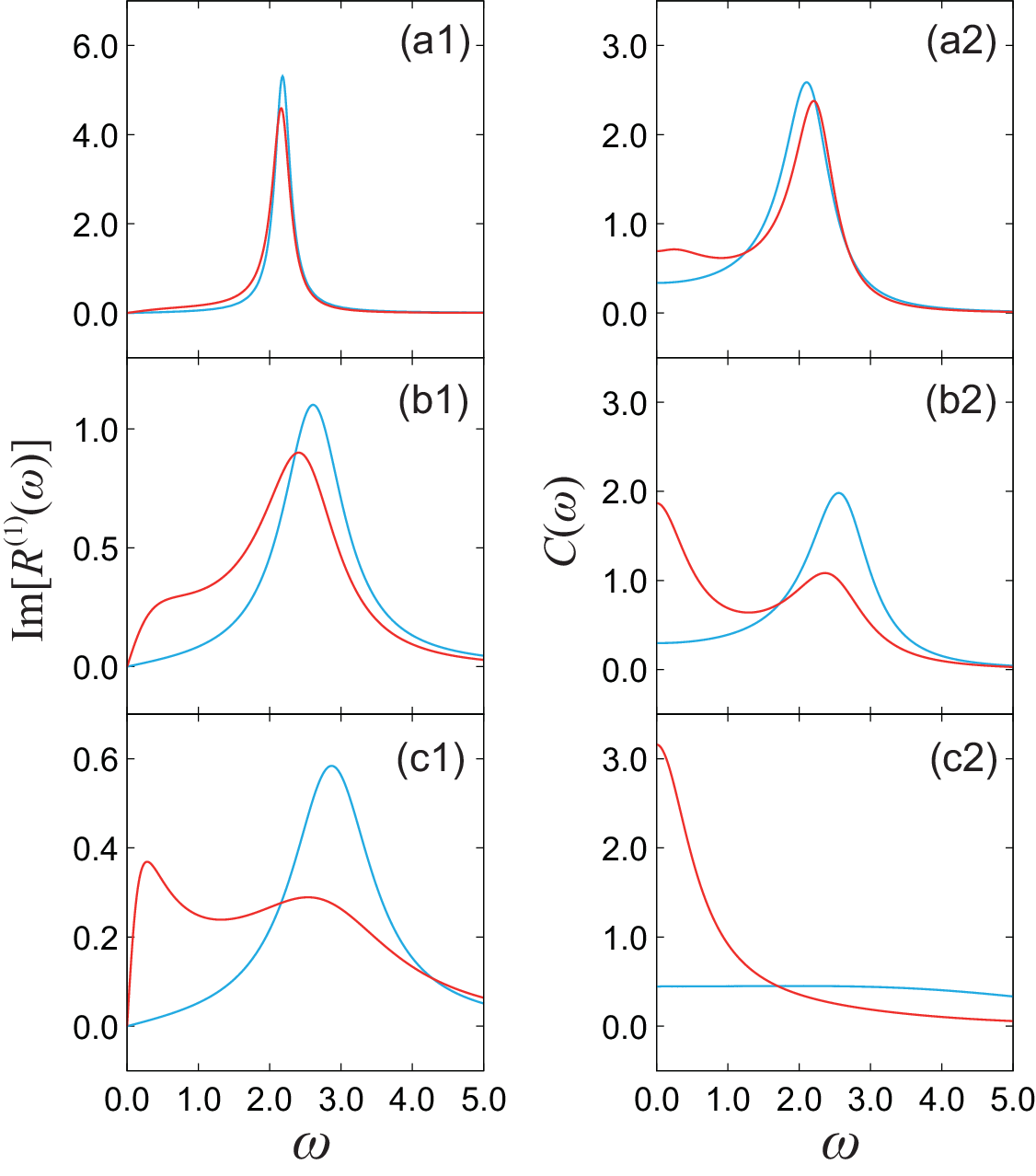}
\caption{\label{fig:LRFCF} Fourier elements of (1) the linear response function and (2) the symmetric correlation function, described as $R^{(1)} [ \omega ]$ and $C [\omega]$, respectively. The red and green curves represent the results from the HEOM-MB and the TCL-Redfield equation, respectively. In the left column, the bath temperature is $\beta = 1.0$ and the SB coupling strengths are (a1) $A = 0.5$ (weak), (b1) $A = 1.0$ (intermediate), and (c1) $A = 1.5$ (strong). In the right column, the SB coupling strength is $A = 1.0$, and the bath temperatures are (a2) $\beta = 10.0$ (cold), (b2) $\beta = 1.0$ (intermediate), and (c2) $\beta = 0.1$ (hot).}
\end{figure}

We depict the imaginary part of the linear response function (LRF) and  the symmetric correlation function in Fig.~\ref{fig:LRFCF}. In the left column, we present the LRF for the cases of (a1) weak  ($A = 0.5)$, (b1) intermediate  ($A = 1.0$), and (c1) strong  ($A = 1.5$) SB coupling. In the weak-coupling case, the HEOM-MB and TCL-Redfield results are almost identical, whereas in the intermediate- and strong-coupling cases, the TCL-Redfield equation result deviates from the HEOM-MB result because the TCL-Redfield equation is a perturbative approach. (See also the results of non-Markovian tests in Ref.~\onlinecite{T15JCP}.)

The symmetric correlation functions are plotted for the cases of (a2) low ($\beta = 10.0$), (b2) intermediate ($\beta = 1.0$), and (c2) high ($\beta = 0.1$) temperature with a SB coupling strength $A = 1.0$. In the low-temperature case, the HEOM-MB and TCL-Redfield results are similar in the resonant frequency region. However, they are different in the low-frequency region, owing to the lack of bath entanglement in the TCL-Redfield result. In the intermediate-temperature case, the HEOM-MB results exhibit two peaks because the subsystem is entangled with the Drude bath mode,\cite{ST02CPL} while the perturbative TCL-Redfield result shows just one peak.  
In the high-temperature case, the HEOM-MB result clearly shows a peak at $\omega=0$ due to bathentanglement, while the TCL-Redfield result shows a featureless flat curve.

\section{Non-Markovian Effects in thermostatic processes}
\label{eq:thermostatic}
\begin{figure}
\includegraphics[width=8cm]{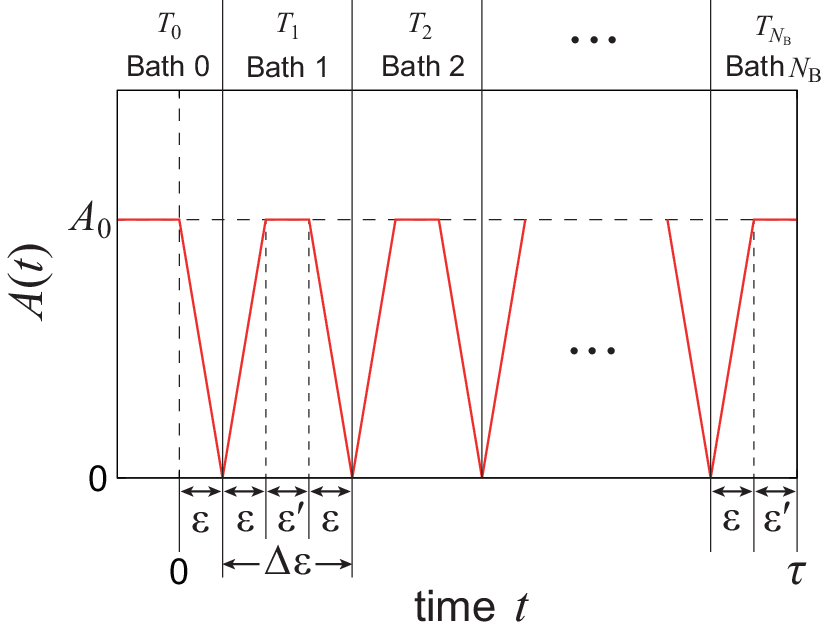}
\caption{\label{fig:discrete} Schematic of the time evolution of $N_{\rm B} + 1$ baths described by $A(t)$. For each bath, 
the periods of the bath attachment, equilibration, and detachment are denoted by $\epsilon$, $\epsilon'$, and $\epsilon$, respectively. The the time step of each bath is then given by $\Delta \epsilon =(2 \epsilon + \epsilon')$.  From Eq.~\eqref{Temp}, the temperature of the $k$th bath ($0\le k \le N_{\rm B}$) is defined as $T_k= T_{TS}(t_k+ \Delta \epsilon /2)$, where $t_k$ is the start time of the $k$th bath.
At time $t = 0$, the system is in the equilibrium state with the $0$th bath.  In the bath attachment and detachment, the SB coupling strength $A ( t )$ respectively increases from $0$ to $A_0$  and decreases from $A_0$ to $0$. We repeat these three processes until the system reaches the equilibration process with the last bath.}
\end{figure}
\begin{figure}
\includegraphics[width=8cm]{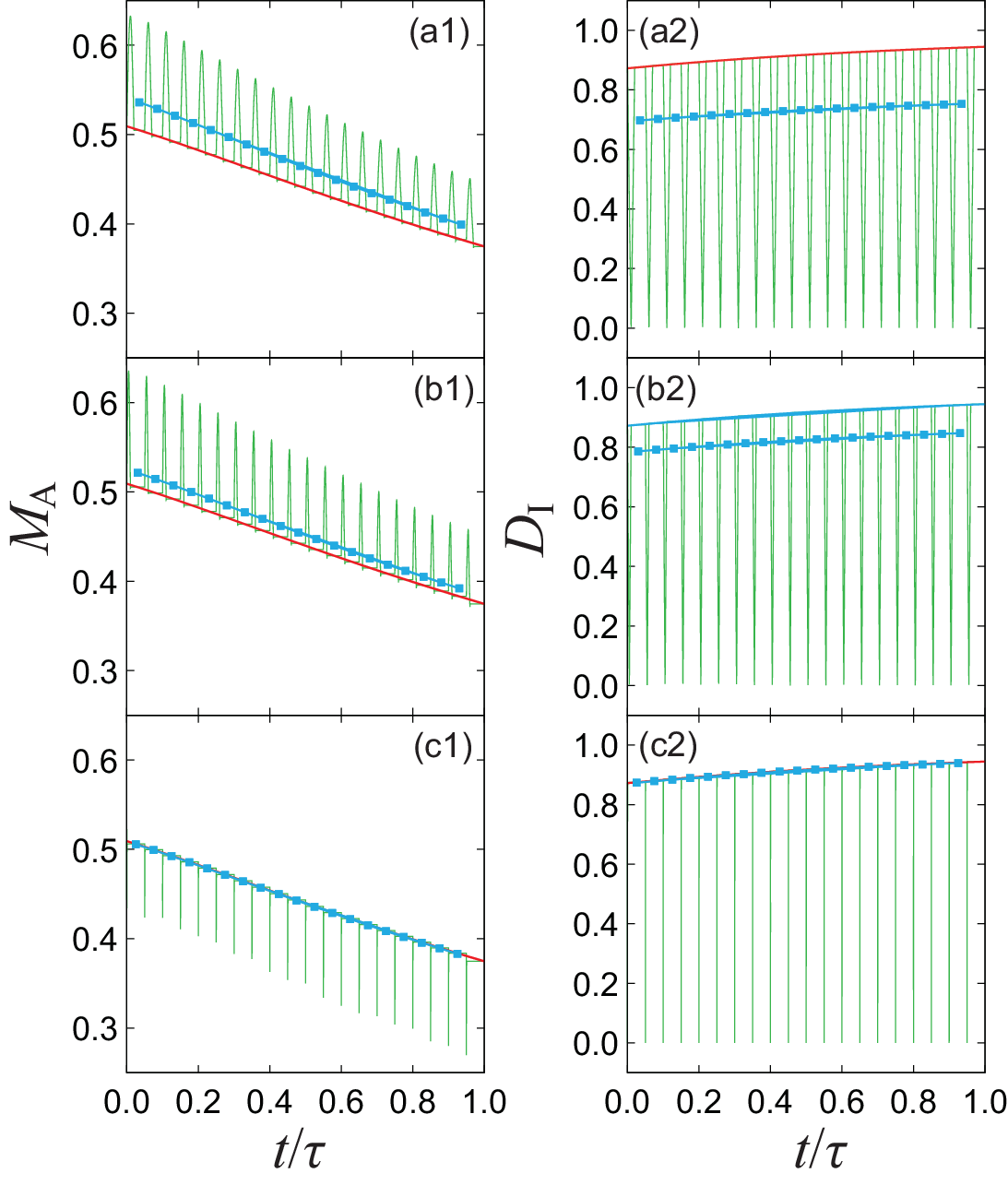}
\caption{\label{fig:Result1} Magnetization $M_{\rm A} ( t )$ and strain $D_{\rm I} ( t )$ as functions of time $t$ for  intermediate coupling strength ($A_0=1.0$) in the cases of (a) slow ($\epsilon = 100$), (b) intermediate ($\epsilon = 50$), and (c) fast ($\epsilon = 0$) bath attachment/detachment. The red and green curves represent the results from the thermostatic bath and stepwise bath models, respectively, and the blue curve with square dots represents the time average for each bath in the stepwise bath model.}
\end{figure}

In the HEOM-MB approach, the thermostatic process is achieved by sequentially changing  heat baths at different temperatures in a stepwise fashion (stepwise bath model). As an approximation, we can also consider only a single bath whose inverse temperature changes as a function of time as $\beta(t)$ (thermostatic bath model) with continuously variation of the hierarchical elements $\hat{\rho}_{\vec{n}} ( t )$.

In the non-Markovian case, the subsystem and the baths are often entangled even in the weak SB case, and so it is not clear that the description of the thermostatic bath model is accurate. Here, we use the HEOM-MB to verify the validity of the thermostatic bath model by comparing the simulated results for the case of a single bath whose inverse temperature changes with time and the case of multiple baths at different temperatures that are sequentially changed in a stepwise fashion. 

We set $B ( t ) = 1.0$ and $E_0 = -1.0$. Then, we consider the following form of temperature change in the thermostatic bath model: 
\begin{equation}
\label{Temp}
T_{\rm TS} ( t ) = 
\begin{cases}
T_{\rm i} & ( t < 0 ),
\\
T_{\rm i} + ( T_{\rm f} - T_{\rm i} ) ( t / \tau ) & ( 0 \leq t < \tau ),
\\
T_{\rm f} & ( \tau \leq t ),
\end{cases}
\end{equation}
where the initial and final temperatures are set to $T_{\rm i} = 1.0$ and $T_{\rm f} = 2.0$.  We then restrict our discussion to the quasi-static case described by $\tau = 1.0 \times 10^4$.  The stepwise bath model is described by  $N_{\rm B} + 1$ baths, with the  time periods of  the bath attachment, equilibration, and detachment steps denoted by $\epsilon$, $ \epsilon'$, and $\epsilon$, respectively (see Fig.~\ref{fig:discrete}).  We repeat these three steps from the zeroth bath to the $N_{\rm B}$th bath until the system reaches the equilibration step of the last bath. The temperature of the stepwise bath model is defined at the midpoint of the time step $\Delta \epsilon \equiv (2 \epsilon + \epsilon')$.
For fixed $\Delta \epsilon$, we change the speeds of the bath attachment/detachment processes described as $\epsilon$ and compare the results with those based on the thermostatic bath model.

In Fig.~\ref{fig:Result1}, we plot the extensive variables  $M_{\rm A} ( t ) $ and $D_{\rm I} ( t )$ as  functions of time as evaluated from Eqs.~\eqref{eq:M} and~\eqref{eq:D}, respectively, in the case of intermediate coupling ($A_0 = 1.0$).\cite{KT22JCP1, KT22JCP2}  We set the number of baths $N_{\rm B} = 20$ in the case of Fig.~\ref{fig:Result1}, so that $\Delta \epsilon= 500$. We then consider three cases with different bath attachment/detachment periods: (a) $\epsilon = 100$ (slow), (b) $\epsilon = 50$ (intermediate), and (c) $\epsilon = 0$ (fast). The red and green curves represent the results from the thermostatic bath model and the stepwise bath model, respectively, and the blue curve with square dots represents the average value of each bath in the stepwise bath model. 
We find that the averaged values from the stepwise bath model deviate from the results obtained from the thermostatic bath model when the switching speed becomes slow, because, for large $\epsilon$, the subsystem follows the dynamics described as $A(t)$ (Fig.~\ref{fig:discrete}) and not as $\beta(t)$: to suppress this deviation, $\epsilon$ must be chosen to be small.

\begin{figure}
\includegraphics[width=8cm]{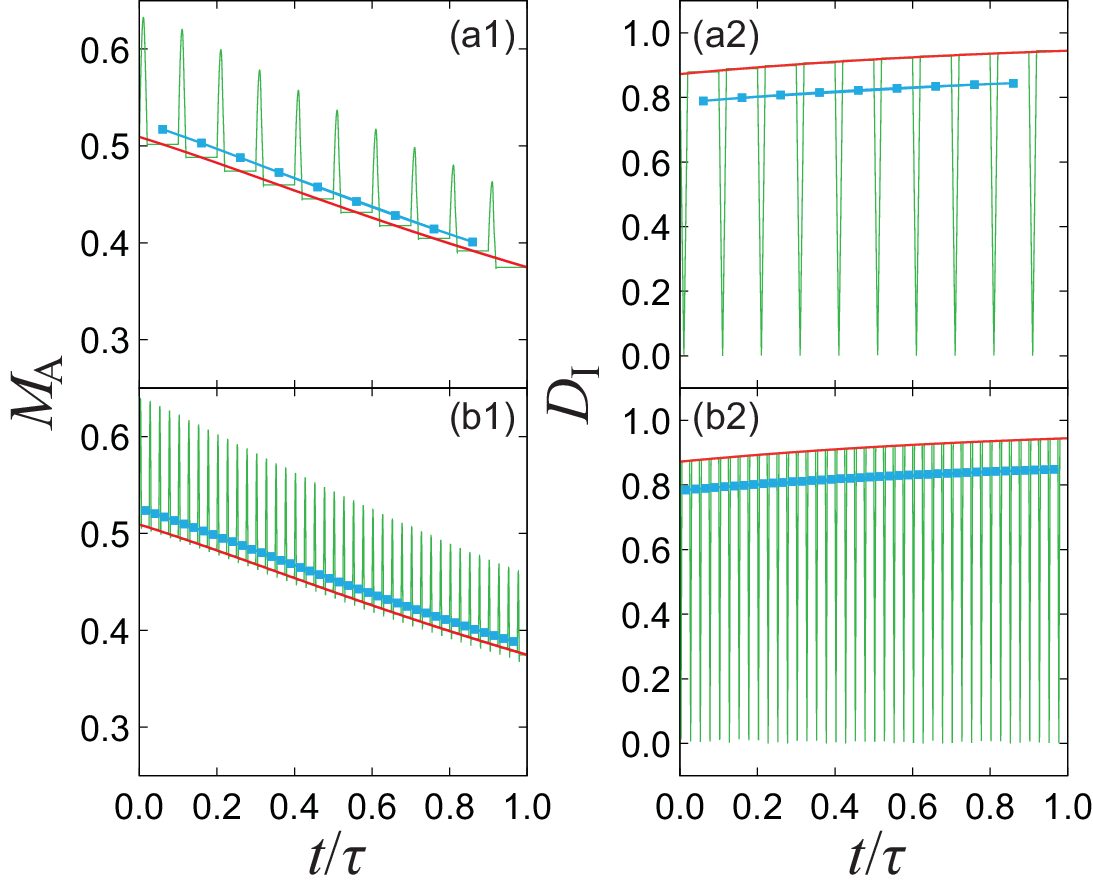}
\caption{\label{fig:Result2} Magnetization $M_{\rm A} ( t )$ and  strain $D_{\rm I} ( t )$ as functions of time $t$  in the intermediate-coupling case ($A_0 = 1.0$)   for  (a) small ($N_{\rm B} = 10$) and (b) large ($N_{\rm B} = 40$) numbers of baths. The red and green curves represent the results from the thermostatic bath and stepwise bath models, respectively, and the blue curve with square dots represents the time average for each bath in the stepwise bath model.}
\end{figure}

In Fig.~\ref{fig:Result2}, we plot $M_{\rm A} ( t )$ and $D_{\rm I} ( t )$ in the intermediate-coupling case ($A_0 = 1.0$) for  (a) small ($N_{\rm B} = 10$) and (b) large ($N_{\rm B} = 40$) numbers of  baths. We fix the ratio $\epsilon / \Delta \epsilon= 0.1$ and set  (a) $\Delta \epsilon= 1000$ with $\epsilon = 100$ and (b) $\Delta \epsilon= 250$ with $\epsilon = 25$. 
From Fig.~\ref{fig:Result2}, we can see that the profiles shown by the blue curves are almost independent of the number of baths $N_{\rm B}$ for  fixed $\epsilon / \Delta \epsilon$.  However, as shown in Table~\ref{table:DLWint}, the entropy production $\Sigma_{\rm tot}$ becomes larger as $N_{\rm B}$ increases, because the bath attachment/detachment steps are not required to be quasi-static processes.

\begin{figure}
\includegraphics[width=8cm]{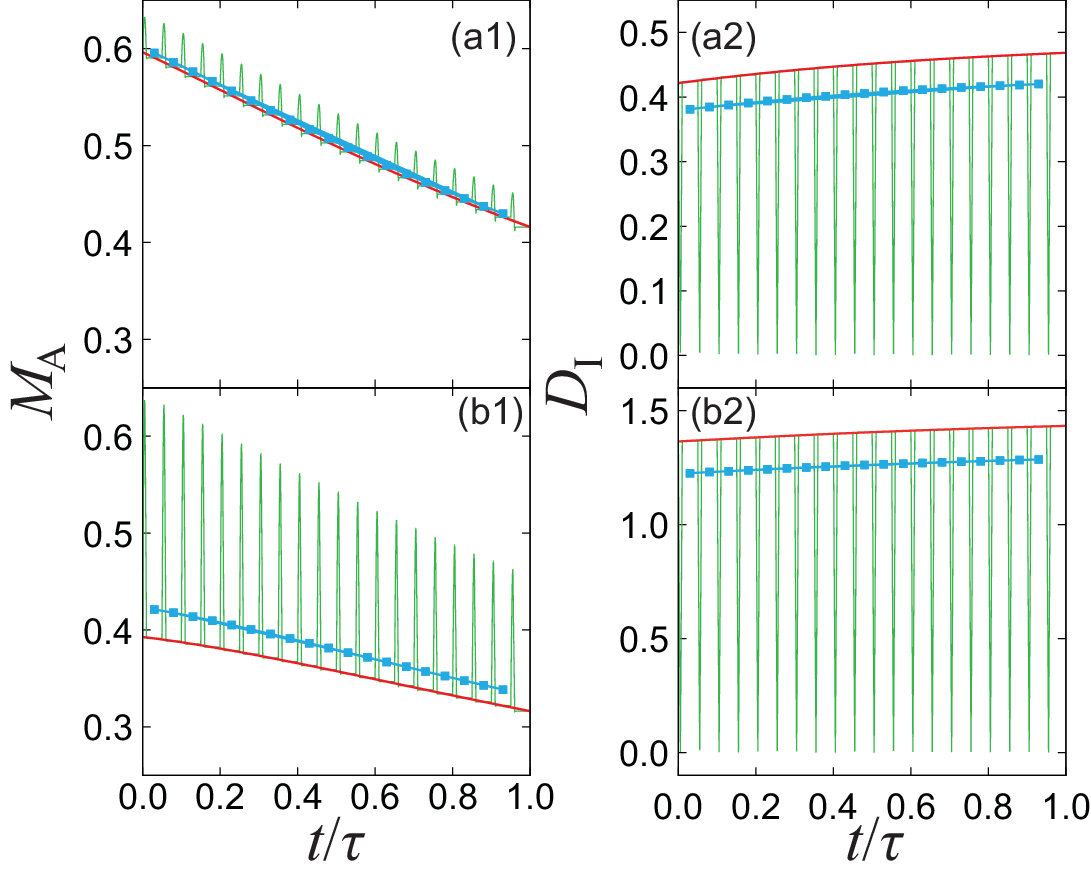}
\caption{\label{fig:Result3} Magnetization $M_{\rm A} ( t )$ and  strain $D_{\rm I} ( t )$ as  functions of time $t$ for  (a) weak ($A_0 = 0.5$) and (b) strong ($A_0 = 1.5$) SB coupling. The red and green curves represent the results from the thermostatic bath and stepwise baths models, respectively. The blue curve with square dots represents the time average for each bath in the stepwise bath model.}
\end{figure}

In Fig.~\ref{fig:Result3}, we plot $M_{\rm A} ( t )$ and $D_{\rm I} ( t )$ for  (a) weak ($A_0 = 0.5$) and (b) strong ($A_0 = 1.5$) SB coupling strength. We set  $N_{\rm B} = 20$,  $\Delta \epsilon= 500$, and $\epsilon = 50$. As the difference between the magnetization at $A ( t ) = 0$ and that at $A ( t ) = A_0$ becomes small for smaller SB coupling strengths (and similarly for the strain), the blue curve approaches the red curve.

\begin{table}

\caption{\label{table:DLWint} Total DL intensive work $\tilde{W}_{\rm tot}^{int}$ and  entropy production $\Sigma_{\rm tot}$ calculated from the stepwise bath model for different numbers of baths, coupling strengths, and time durations.}
\begin{ruledtabular}
\begin{tabular}{cccccc}
\hspace{0.3cm} $N_{\rm B}$ \hspace{0.3cm} & \hspace{0.3cm} $A_0$ \hspace{0.3cm} 
& \hspace{0.3cm} $\epsilon$ \hspace{0.3cm} & \hspace{0.3cm} $\epsilon'$ \hspace{0.3cm} 
& \hspace{0.3cm} $\tilde{W}_{\rm tot}^{int}$ \hspace{0.3cm} & $\Sigma_{\rm tot}$
\\
\hline \hline
$10$ & $1.0$ & $100$ & $800$ & $0.841$ & $0.102$
\\
$20$ & $0.5$ & $50$ & $400$ & $0.675$ & $8.17 \times 10^{-2}$
\\
$20$ & $1.0$ & $100$ & $300$ & $0.926$ & $0.188$
\\
$20$ & $1.0$ & $50$ & $400$ & $1.10$ & $0.358$
\\
$20$ & $1.0$ & $0$ & $500$ & $14.0$ & $13.3$
\\
$20$ & $1.5$ & $50$ & $400$ & $1.88$ & $0.884$
\\
$40$ & $1.0$ & $25$ & $200$ & $2.06$ & $1.32$
\\
$100$ & $0.5$ & $500$ & $0$ & $0.638$ & $4.43 \times 10^{-2}$
\end{tabular}
\end{ruledtabular}
\end{table}

We have summarized the results of Figs.~\ref{fig:Result1}--\ref{fig:Result3}  in terms of the DL total intensive work $\tilde{W}_{\rm tot}^{int}$ and the entropy production $\Sigma_{\rm tot}$, along with the bath parameters, in Table~\ref{table:DLWint}.
For the stepwise bath model, the entropy production is evaluated from Eq.~\eqref{eq:loss}, with the change of Planck potential being evaluated from the Gibbs energy using Eq.~\eqref{eq:Planck-Gibbs} (see Appendix~\ref{sec:ComputePlanck}). 
Note that the work that is wasted because the system does not change quasi-statically is the entropy production.\cite{KT24JCP2}  
Since the process described as a thermostatic bath model is chosen to be quasi-static, the value of $\Sigma_{\rm tot}$ indicates the deviation from the quasi-static process in the case of the stepwise model;  $\Sigma_{\rm tot}$ becomes smaller as $\epsilon$ becomes larger  (e.g., $\epsilon=500$)
because the bath attachment/detachment process becomes slow and the equilibration process is effective.
In addition, for shorter $\epsilon$, the magnetization and strain results from the thermostatic model and the time-averaged results from the stepwise model are in better agreement. Thus, for a valid description of the thermostatic bath model to be obtained, the value of $\epsilon$ must be adjusted to an appropriate value that is neither too long nor too short.

It is also found that $\Sigma_{\rm tot}$ becomes large as the SB coupling strength $A_0$ increases. 
In this case,  $\tau$ must be set to a large value to ensure that the thermostatic bath model is an approximation of the stepwise bath model.

\section{Quantum Carnot Cycle}
\label{sec:Carnot}

\begin{figure}

\includegraphics[width=8cm]{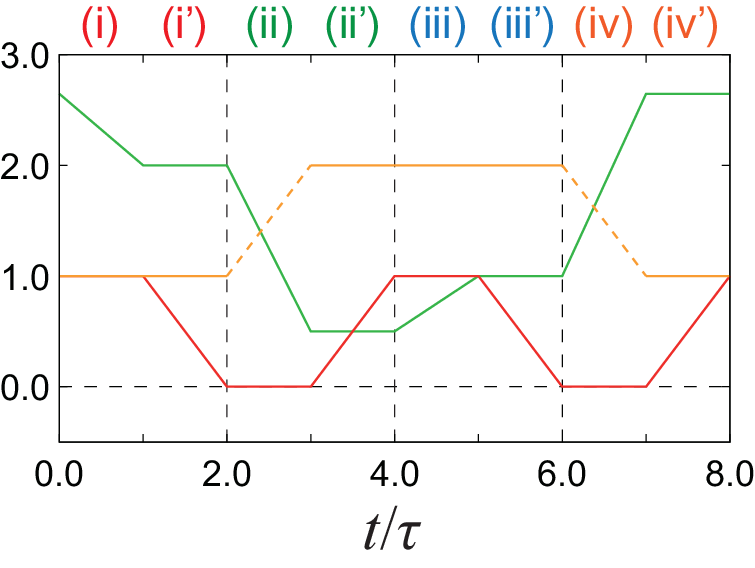}
\caption{\label{fig:Parameter} Time profiles of $B ( t )$ (green), $A ( t )$ (red), and $\beta ( t )$ (orange).}

\end{figure}

Previously, using the same model as presented in Sec.~\ref{sec:modelHamiltonian}, we have studied the efficiency of Carnot cycles\cite{KT22JCP2} by evaluating the Gibbs energy $G_{\rm tot}^{\rm qst}(t)$ as the quasi-static intensive work $W^{int}_{\rm tot}(t)$  on the basis of the minimum work principle 
[Kelvin--Planck statement: $W^{int}_{\rm tot} ( t ) \geq \Delta G_{\rm tot}^{\rm qst}( t )$].\cite{ST20JCP, KT22JCP1} 
 Although the results are essentially the same, here we repeat the simulation based on the DL minimum work principle expressed in terms of the intensive and extensive variables,\cite{KT24JCP1} using the HEOM-MB. Although calculations based on this theory have been performed for an anharmonic Brownian system,\cite{KT24JCP1,KT24JCP2,KT24JCP3} the treatment is not the same for spin--boson systems, since the SB interaction is not included in the heat bath. 

The quantum Carnot cycle consists of the following eight steps: (i) hot isotherm, (i$'$) hot bath detachment, (ii) adiabatic from hot to cold, (ii$'$) cold bath attachment, (iii) cold isotherm, (iii$'$) cold bath detachment, (iv) adiabatic from cold to hot, and (iv$'$) hot bath attachment. We assume that the duration of each step is the same and denote it by $\tau$. The inverse temperatures of the hot and cold baths are $\beta_{\rm H} = 1.0$ and $\beta_{\rm C} = 2.0$, respectively. We fix $E ( t ) = 1.0$ and set the time profiles of $B ( t ), A ( t )$, and $\beta ( t )$ as presented in Fig.~\ref{fig:Parameter}. 

Adiabatic processes are introduced in this model using the adiabatic condition defined as\cite{KT22JCP2}
\begin{eqnarray}
d H^{\rm qst} = - M_{\rm A}^{\rm qst} d B^{\rm qst} .
\end{eqnarray}
Assuming that the system density operator is the Boltzmann distribution of the system Hamiltonian, $\hat{\rho}_{\rm A} = e^{- \beta \hat{H}_{\rm A}} / \tr_{\rm A} \{ e^{- \beta \hat{H}_{\rm A}} \}$ during the adiabatic process and  that the inverse temperatures of the Boltzmann distribution at the beginning and end of the adiabatic process are $\beta_{\rm H}$ (or $\beta_{\rm C}$) and $\beta_{\rm C}$ ( or $\beta_{\rm H}$ ), respectively, we obtain the following conditions for the two adiabatic processes:\cite{Kosloff_2020,KT22JCP2}
\begin{eqnarray}
\beta_{\rm H} \sqrt{E_0^2 + B^2 ( 2 \tau )} = \beta_{\rm C} \sqrt{E_0^2 + B^2 ( 3 \tau )} 
\end{eqnarray}
and
\begin{eqnarray}
\beta_{\rm C} \sqrt{E_0^2 + B^2 ( 6 \tau )} = \beta_{\rm H} \sqrt{E_0^2 + B^2 ( 7 \tau )}.
\end{eqnarray}
In Fig.~\ref{fig:Parameter}, we choose $B ( 3 \tau ) = 0.5$ and $B ( 6 \tau ) = 1.0$, and so we obtain $B ( 2 \tau ) = 2.0$ and $B ( 7 \tau ) = \sqrt{7.0}$.

\begin{figure}
\includegraphics[width=8cm]{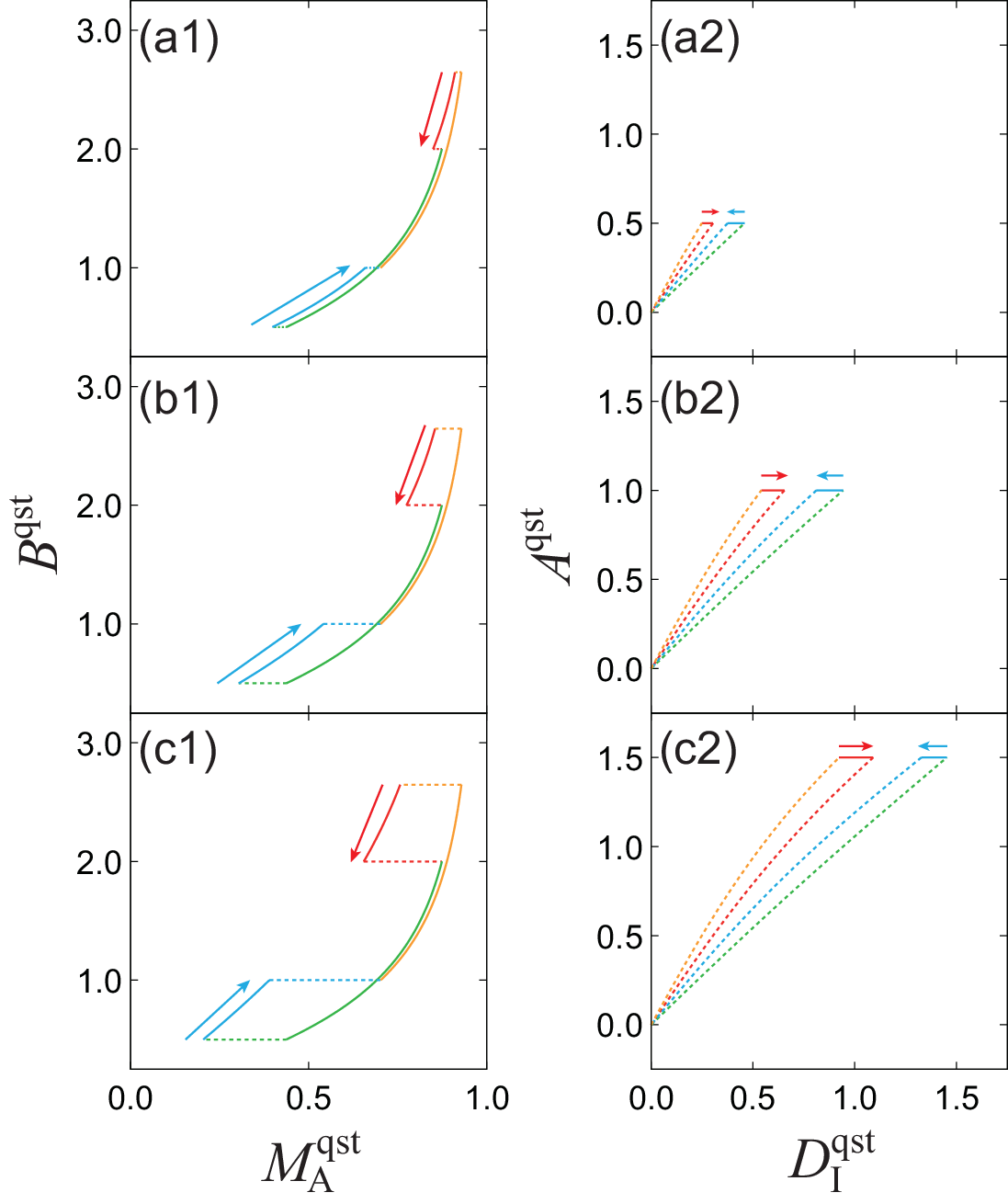}
\caption{\label{fig:BMAD}  (1) $B^{\rm qst}$--$M_{\rm A}^{\rm qst}$ diagrams and (2) $A^{\rm qst}$--$D_{\rm I}^{\rm qst}$ diagrams for the quasi-static Carnot cycle for  (a) weak ($A_0 = 0.5$), (b) intermediate ($A_0 = 1.0$), and (c) strong ($A_0 = 1.5$) SB coupling strength. The red, green, blue, and orange solid curves represent the hot isothermal, hot to cold adiabatic, cold isothermal, and cold to hot adiabatic processes, respectively, while the red, green, blue, and orange dashed curves represent the hot bath detachment, cold bath attachment, cold bath detachment, and hot bath attachment processes, respectively. The  hot isothermal step (i) starts from the red arrow, and the cold isothermal step (iii)  starts from the blue arrow.}
\end{figure}

In Fig.~\ref{fig:BMAD},  $B^{\rm qst}$--$M_{\rm A}^{\rm qst}$ and $A^{\rm qst}$--$D_{\rm I}^{\rm qst}$ diagrams are plotted for cases of (a) weak $A_0 = 0.5$, (b) intermediate $A_0 = 1.0$, and (c) strong $A_0 = 1.5$ coupling.  The area enclosed by the counterclockwise and clockwise curves in the diagrams represents the work done by the system and the work done on the system, respectively.

\begin{figure}
\includegraphics[width=4.5cm]{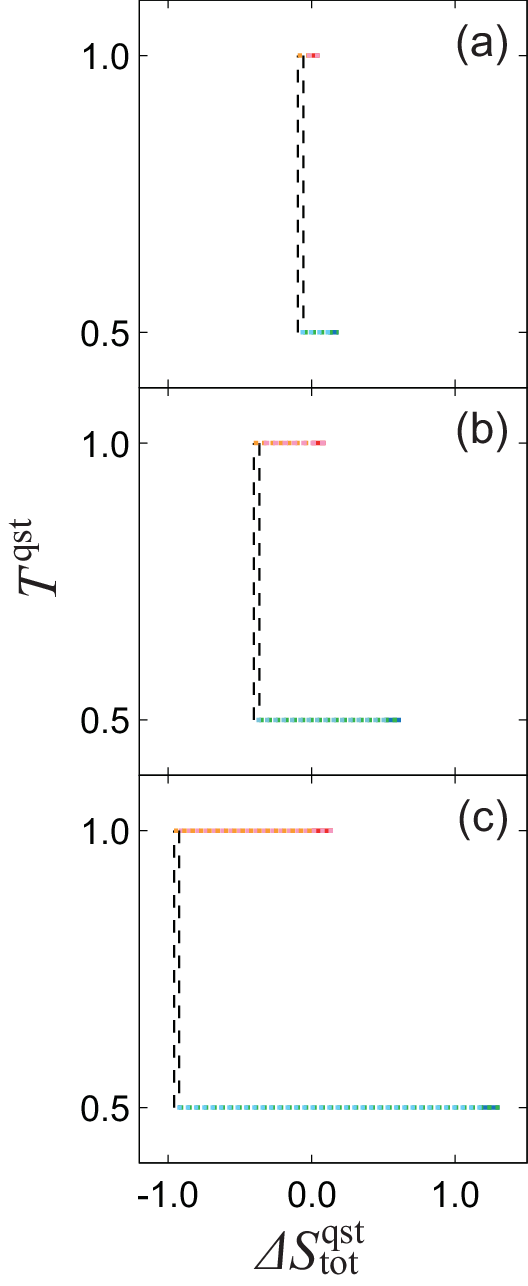}
\caption{\label{fig:TS}  $T^{\rm qst}$--$\Delta S_{\rm tot}^{\rm qst}$ diagrams  for  (a) weak ($A_0 = 0.5$), (b) intermediate ($A_0 = 1.0$), and (c) strong ($A = 1.5$) SB coupling. The red and blue solid lines represent the hot isothermal and cold isothermal processes, respectively, while the dashed red, green, blue, and orange lines represent the hot bath detachment, cold bath attachment, cold bath detachment, and hot bath attachment processes, respectively. The black dashed lines represent the adiabatic processes.}
\end{figure}

We plot  $T^{\rm qst}$--$\Delta S_{\rm tot}^{\rm qst}$ diagrams in Fig.~\ref{fig:TS} for cases of (a) weak $A_0 = 0.5$, (b) intermediate $A_0 = 1.0$, and (c) strong $A_0 = 1.5$ SB coupling. In the adiabatic process, we assume that the entropy is constant, because the system does not exchange heat with the heat bath during the process, and we denote this process by the black dashed lines in Fig.~\ref{fig:TS}.

Although the definitions of the intensive variables are different, this result is the same as the previous result where the Gibbs energy was used to define intensive variables.
However, the intensive quantities defined in this paper can also describe non-equilibrium states, and by including the entropy production, the work diagram can  be employed to analyze non-equilibrium processes as well.\cite{KT24JCP2} Exploration in this direction is left to a future study.

\section{Remarks}
\label{sec:conclude} 

Both the anharmonic Brownian and the spin--boson models are widely used and  have been extensively studied in non-equilibrium statistical physics. In this paper, we have extended the spin--boson model by introducing multiple Drude baths to derive the HEOM-MB, whereas in our previous paper, we considered an anharmonic potential system coupled with multiple Ohmic baths to derive the T-QFPE. While the HEOM-MB treats the SB interaction as separate from the subsystem,\cite{KT15JCP,KT16JCP,ST21JPSJ,ST20JCP,KT22JCP1,KT22JCP2} the T-QFPE treats it as part of a heat bath  thanks to the counterterm.\cite{KT24JCP1,KT24JCP2,KT24JCP3} Note that in the spin--boson-based model, the inclusion of a counterterm, as in the T-QFPE case, is not possible, since the Hamiltonian of the subsystem and that of the SB interaction are commutative.

The anharmonic Brownian model with an Ohmic bath is equivalent to the Markovian Langevin description in the classical limit and fits well with the description of classical thermodynamics, which does not explicitly consider the interaction with the bath. On the other hand, the spin--boson model is purely quantum and non-Markovian, with  the Markovian limit of the system exhibiting anomalous behavior (Appendix~\ref{sec:OhmicLimit}).  Furthermore, the equilibrium state of the subsystem is not its own Boltzmann distribution and cannot be factorized from the bath, owing to the bathentanglement. 

These facts contradict the Markovian assumption for stochastic thermodynamics\cite{Schmiedl_2008,Stochastic_Seifert_2012,Esposito2011,strasberg2022quantum,RaoEsposit2018,PhysRevE.105.064124} and the factorization assumption for the fluctuation theorem,\cite{RevModPhys.81.1665,RevModPhys.83.771,sagawa20232law} and they need to be reconsidered through the use of the HEOM-MB and T-QFPE: such theories should be derived from a general theory that covers both classical and quantum dynamics,  even in the non-Markovian case.
The equilibrium and nonequilibrium thermodynamic theories based on the DL minimum work principle derived from the T-QFPE\cite{KT24JCP1,KT24JCP2} and the present HEOM-MB make such exploration worthwhile.

The source code for the spin--boson system provided in this paper and that for the anharmonic Brownian system in the previous paper,\cite{KT24JCP3} which can take the classical limit, were developed to enable researchers to perform thermodynamic numerical experiments without being experts in open quantum dynamics theory. Extensions to non-equilibrium systems are also possible.\cite{KT24JCP2}  Although both codes have been demonstrated for simple systems, they can be extended to study electron transfer systems,\cite{TT09JPSJ,TT10JCP,Shi2009ET, IshizakiPhuc2018} exciton transfer systems,\cite{Schuten09,FujihashiIshi2015,Hein_2012} spin lattice systems,\cite{NT18PRA,NT21BJCP} and Holstein--Peierls\cite{CT21JCP} and Holstein--Hubbard\cite{NT21JCP} models.
Moreover the DL minimum work principle,\cite{KT24JCP1,KT24JCP2} allows us to bring a thermodynamic perspective from the quasi-static case to the highly non-equilibrium case for the exploration of various systems, including those mentioned above.

\section*{Supplementary Material}
Numerical integration codes for the HEOM-MB and four demo codes (three for non-Markovian tests and one for the quantum Carnot engine) are provided as supplementary material, where the manual for these codes can be found in the ReadMe.pdf file.

\section*{Acknowledgments}
Y. T. would like to thank Massimiliano Esposito for the stimulating discussions and Jianshu Cao for providing the references for the anomaly of the spin--boson system with Ohmic SDF. Y.T. was supported by JSPS KAKENHI (Grant No.~B21H01884). S.K. acknowledges was supported by Grant-in-Aid for JSPS Fellows (Grant No. 24KJ1373).

\section*{Author declarations}
\subsection*{Conflict of Interest}
The authors have no conflicts to disclose.

\section*{Author Contributions}
{\bf Shoki Koyanagi}: Investigation (equal); Software (lead); Writing – original draft (lead). {\bf Yoshitaka Tanimura}: Conceptualization (lead); Investigation (equal); Writing – review and editing (lead).

\section*{Data availability}
The data that support the findings of this study are available from the corresponding author upon reasonable request.

\appendix

\section{The change in the bath energy}
\label{sec:BathEnergy}
From the energy conservation law of the total system, the change in the bath energy is expressed as 
\begin{eqnarray}
\label{eq:EnergyConservation}
-\frac{\partial}{\partial t} H_{\rm B} ( t )
= \frac{\partial}{\partial t} ( H_{\rm A} ( t ) + H_{\rm I} ( t ) ) - \frac{d W_{\rm tot}^{int} ( t )}{d t}.
\end{eqnarray}
Replacing the right-hand side of Eq.~\eqref{eq:EnergyConservation} by Eqs.~\eqref{eq:DefHA},~\eqref{eq:InteractionEnergy}, and~\eqref{eq:DefWint-tot} and using the HEOM [Eq.~\eqref{ModelHEOM}], we obtain Eq.~\eqref{eq:expectEB}.\cite{KT16JCP,KT22JCP2}

\section{Enthalpy from the SB interaction}
\label{sec:TDEnthalpy}

On the basis of our previous work on the Kelvin--Planck cycle\cite{ST20JCP} and Carnot cycle,\cite{KT22JCP1} the Boltzmann enthalpy can be introduced as $H_{\rm A}^{\rm Boltz} ( t ) = \partial ( \beta G_{\rm A}^{\rm qst} ( t ) ) / \partial \beta$. However, $H_{\rm A}^{\rm Boltz} ( t )$ is different from 
$H_{\rm A}^{\rm qst} ( t ) = \tr_{\rm A} \{ \hat{H}_{\rm A} ( t ) \hat{\rho}_{\rm A}^{\rm qst} ( t ) \}$ defined as in Eq.~\eqref{eq:DefHA}, because the Boltzmann enthalpy includes the contribution from the system part of the SB interaction,\cite{ST20JCP} in contrast to the Ohmic Brownian case.\cite{KT24JCP1,KT24JCP2,KT24JCP3}   Here, we show that this difference depends on the form of the SDF. To illustrate this point, we consider here  the case of a single bath [$N_{\rm B} = 0$ in Eq.~\eqref{eq:Htotal} and omitting the bath index $k$].  Then, we introduce the partition function defined as $Z_{\rm A} = \tr_{\rm tot} \{ e^{- \beta \hat{H}_{\rm tot}} \} / \tr_{\rm B} \{ e^{- \beta \hat{H}_{\rm B}} \}$, where $\beta$ is the inverse temperature of the total system. Because $G_{\rm A}^{\rm qst} = - \beta^{-1} \ln Z_{\rm A}$,\cite{ST20JCP,KT22JCP1} the enthalpy is expressed as $H_{\rm A}^{\rm Boltz} = - \partial ( \ln Z_{\rm A} ) / \partial \beta$. For the SB Hamiltonian, the partition function is expressed as 
\begin{eqnarray}
\label{eq:PartitionFunction1}
Z_{\rm A} = \tr_{\rm A} \{ e^{- \beta \hat{H}_{\rm A}} \hat{\mathcal{F}}_{\rm IF} ( \hat{V} ( \tau ) ; \beta \hbar ) \} ,
\end{eqnarray}
where the influence functional is defined as\cite{T14JCP,T15JCP}
\begin{equation}
\label{eq:InfluenceFunctional1}
\begin{split}
\hat{\mathcal{F}}_{\rm IF} ( \hat{V} ( \tau ) ; \beta \hbar )
&= \mathcal{T} \exp \left[ \frac{A^2}{\hbar^2} \int^{\beta \hbar}_0 d \tau \int^{\tau}_0 d \tau' 
\hat{V} ( \tau ) \hat{V} ( \tau' ) \right. \\
& 
\times \left.
\int^\infty_0 J ( \omega ) 
\frac{\cosh \left( \beta \hbar \omega / 2 - \omega ( \tau - \tau' ) \right)}
{\sinh \left( \beta \hbar \omega / 2 \right)} d \omega \right] .
\end{split}
\end{equation}
Here, $\hat{V} ( \tau ) = e^{\tau \hat{H}_{\rm A}} \hat{V} e^{- \tau \hat{H}_{\rm A}}$ is the interaction picture of the system operator $\hat{V}$.  The $\beta$ derivative of $Z_{\rm A} $ is expressed as $H_{\rm A}^{\rm Boltz} =H_{\rm A} + H_{\rm A}'$, where $H_{\rm A} = \tr_{\rm A} \{ \hat{H}_{\rm A} e^{- \beta \hat{H}_{\rm A}} \hat{\mathcal{F}}_{\rm IF} \}/ Z_{\rm A}$ and  $H_{\rm A}' =- \tr_{\rm A} \{ e^{- \beta \hat{H}_{\rm A}} ( \partial \hat{\mathcal{F}}_{\rm IF} / \partial \beta ) \}/ Z_{\rm A}$. The latter corresponds to the enthalpy from the system part of the SB interaction.

To clarify the $\beta$ dependence, we rewrite Eq.~\eqref{eq:InfluenceFunctional1} using the dimensionless variables $\tilde{\omega} = \beta \hbar \omega$, $\bar{\tau} = \tau / \beta \hbar$, and $\bar{\tau}' = \tau' / \beta \hbar$ as
\begin{equation}
\label{eq:AppD2}
\begin{split}
&\mathcal{F}_{\rm IF} ( \hat{V} ( \tau ) ; \beta \hbar )
= \mathcal{T} \exp \left[ A^2 \int^1_0 d \bar{\tau} \int^{\bar{\tau}}_0 d \bar{\tau}' 
\hat{V} ( \beta \hbar \bar{\tau} ) \hat{V} ( \beta \hbar \bar{\tau}' ) \right. \\
&
\qquad \times \left.
\int^\infty_0 \beta \hbar J \left( \frac{\tilde{\omega}}{\beta \hbar} \right)
\frac{\cosh ( \tilde{\omega} / 2 - \tilde{\omega} ( \bar{\tau} - \bar{\tau}' ) )}
{\sinh ( \tilde{\omega} / 2 )} d \tilde{\omega} \right] .
\end{split} 
\end{equation}
The $\beta$ derivative of the above equation consists of terms obtained from the derivatives of $\hat{V} ( \beta \hbar \bar{\tau} ) \hat{V} ( \beta \hbar \bar{\tau}' ) $ and $\beta \hbar J ( \tilde{\omega} / \beta \hbar )$.  The first contribution disappears because it is expressed as $\tr_{\rm tot} \{ [ \hat{H}_{\rm A} , \hat{H}_{\rm I} ] \hat{\rho}_{\rm tot}^{\rm eq} \} / 2$, where $\hat{\rho}_{\rm tot}^{\rm eq} = e^{- \beta \hat{H}_{\rm tot}} / \tr_{\rm tot} \{ e^{- \beta \hat{H}_{\rm tot}} \}$ is the equilibrium total density operator, and because of the relation $[ \hat{H}_{\rm A} , \hat{H}_{\rm I} ] = [ \hat{H}_{\rm A} , \hat{H}_{\rm tot} ]$. 
Thus, $H_{\rm A}'$ is evaluated from the term involving the $\beta$ derivative of  $\beta \hbar J ( \tilde{\omega} / \beta \hbar )$. 
In the Ohmic SDF case, $\beta \hbar J ( \tilde{\omega} / \beta \hbar )$ is independent of $\beta$, and we have $H_{\rm A}^{\rm Boltz} ( t ) = H_{\rm A}^{\rm qst} ( t )$, as we discussed in the Ohmic Brownian case.\cite{KT24JCP1,KT24JCP3}  
In the Drude SDF case, by introducing the DL inverse of the noise correlation time, $\tilde{\gamma} \equiv \beta \hbar \gamma$, we can make $\beta \hbar J ( \tilde{\omega} / \beta \hbar )$ temperature-independent.  
Thus, the enthalpy of subsystem without the contribution of the SB interaction is expressed as
\begin{eqnarray}
\label{eq:KineticEnthalpy1}
H_{\rm A}^{\rm qst} ( t ) = \left. \frac{\partial}{\partial \beta} ( \beta G_{\rm A}^{\rm qst} ( t ) )
\right|_{B , A , \tilde{\gamma}} .
\end{eqnarray}
The Planck potential for $A^{\rm qst} ( t ) \neq 0$ is evaluated from Eq.~\eqref{eq:Planck-Gibbs} and expressed as Eq.~\eqref{eq:Planck-U}.
By changing the fixed variable from $\tilde{\gamma}$ to $\gamma$, Eq.~\eqref{eq:KineticEnthalpy1} is rewritten as
\begin{eqnarray}
\label{eq:KineticEnthalpy2}
H_{\rm A}^{\rm qst} ( t ) = H_{\rm A}^{\rm Boltz} ( t ) + \gamma \alpha^{\rm qst} ( t ) ,
\end{eqnarray}
where $\alpha^{\rm qst} ( t ) \equiv - ( \partial G_{\rm A}^{\rm qst} ( t ) / \partial \gamma ) |_{\beta , B , A}$ is regarded as the non-Markovian bath susceptibility.  From Eq.~\eqref{eq:KineticEnthalpy2}, the system part of the SB interaction is expressed as ${H_{\rm A}'}^{\rm qst} ( t ) = \gamma \alpha^{\rm qst} ( t )$.

Since the Gibbs energy is an extensive variable, 
$\alpha^{\rm qst}  (t)$ is also an extensive variable, whereas $\gamma$ is an intensive variable, the value of which is independent of the system size. Thus, we can regard Eq.~\eqref{eq:KineticEnthalpy2} as the Legendre transformation between $H_{\rm A}^{\rm qst} ( t )$ and $H_{\rm A}^{\rm Boltz} ( t )$. 

\begin{figure}
\includegraphics[scale=0.5]{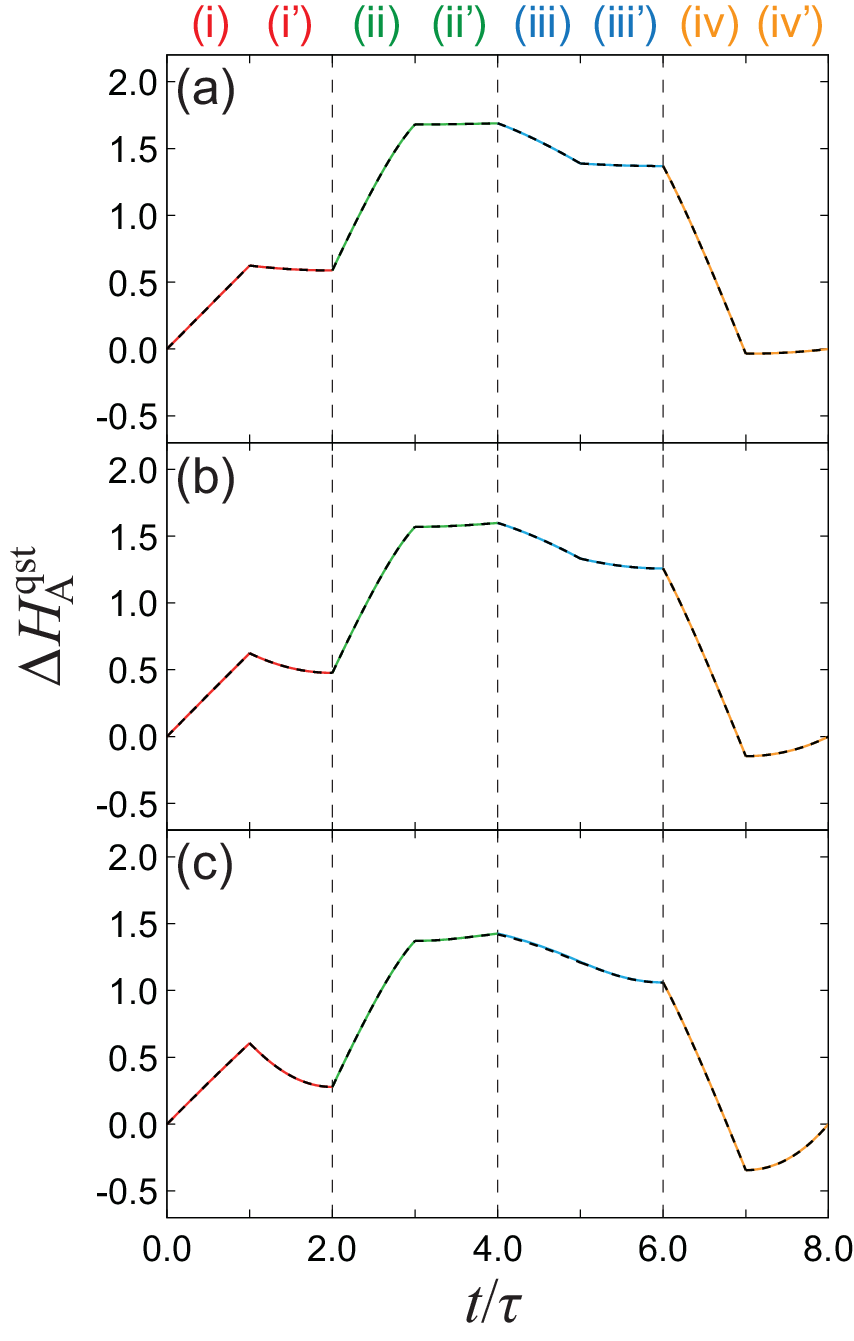}
\caption{\label{fig:H}Quasi-static enthalpy for the quantum Carnot cycle in the cases of  (a) weak ($A = 0.5$), (b) intermediate ($A = 1.0$), and (c) strong ($A = 1.5$) SB coupling. The colored and black dashed curves represent the enthalpy computed from the quasi-static Gibbs energy [the right-hand side of Eq.~\eqref{eq:KineticEnthalpy1}] and the expectation value of the subsystem Hamiltonian $\tr_{\rm A} \{ \hat{H}_{\rm A} ( t ) \hat{\rho}_{\rm A} ( t ) \}$, respectively. The colored  and  black dashed curves are almost overlapping.}
\end{figure}

To verify the relation in Eq.~\eqref{eq:KineticEnthalpy1}, we numerically calculate  the right-hand side of Eq.~\eqref{eq:KineticEnthalpy1} for the quantum Carnot cycle. The parameters are the same as those in Sec.~\ref{sec:Carnot}. We plot the result in Fig.~\ref{fig:H} for the cases of (a) weak ($A = 0.5$), (b) intermediate ($A = 1.0$), and (c) strong ($A = 1.5$) SB coupling. The colored  and the black dashed curves represent $\partial (\beta G_{\rm A}^{\rm qst} ) / \partial \beta |_{B , A , \tilde{\gamma}}$ and $\tr_{\rm A} \{ \hat{H}_{\rm A} ( t ) \hat{\rho}_{\rm A}^{\rm qst} ( t ) \}$, respectively. Regardless of the SB coupling strength, the two curves overlap, and thus Eq.~\eqref{eq:KineticEnthalpy1} is verified.

\section{Numerical implementation of the HEOM-MB}
\label{sec:RKF}

\begin{table}

\caption{\label{table:NK} Truncation numbers $N$ and $K$ used to perform the benchmark calculations for the various parameters $\beta$ and $A$.}
\begin{ruledtabular}
\begin{tabular}{cccc}
\hspace{0.7cm} $\beta$ \hspace{0.7cm} & \hspace{0.7cm} $A$ \hspace{0.7cm} 
& \hspace{0.7cm} $N$ \hspace{0.7cm} & \hspace{0.7cm} $K$ \hspace{0.7cm}
\\
\hline
$0.1$ & $1.0$ & $25$ & $1$
\\
$1.0$ & $0.5, 1.0$ & $8$ & $4$
\\
$1.0$ & $1.5$ & $9$ & $4$
\\
$10.0$ & $1.0$ & $8$ & $6$
\end{tabular}
\end{ruledtabular}

\end{table}

The choice of time step is important when performing numerical integration. 
The program code provided in the supplementary material of this paper uses the Runge--Kutta--Fehlberg (RKF) method\cite{fehlberg1969low} for integrating Eq.~\eqref{ModelHEOM} to adaptively select the time step size to simplify its use.  When using the RKF, an appropriate time step $\delta t$ is chosen such that the estimated error $\epsilon_{\rm est}$ is less than the desired tolerance $\epsilon_{\rm TOL}$.  

The estimation of the error in the time evolution from time $t$ to $\delta t$ is performed as follows. First, we compute the ADOs at time $t + \delta t$ using the fourth- and fifth-order Runge--Kutta methods, and we write the ADOs as $\hat{\rho}_{\vec{n}}^{(4)} ( t + \delta t )$ and $\hat{\rho}_{\vec{n}}^{(5)} ( t + \delta t )$.
Here, we have found that two ADOs [$\hat{\rho}_{\vec{0}} ( t )$ and $\hat{\rho}_{N \vec{e}_K} ( t )$] are sufficient for the error estimation, because the HEOM-BM for $\hat{\rho}_{N \vec{e}_K} ( t )$ includes the largest damping term $\sum_{j = 0}^K n_j \nu_j ( t )$, which becomes the main source of the numerical error.  Then, using the two ADOs at time $t + \delta t$, we define the estimated error $\epsilon_{\rm est}$ as 
the maximum of $| \langle + | ( \hat{\rho}_{\vec{0}}^{( 4 )} - \hat{\rho}_{\vec{0}}^{( 5 )} ) | + \rangle |$
and $| \mathrm{Re} \langle + | ( \hat{\rho}_{\vec{n}}^{( 4 )} - \hat{\rho}_{\vec{n}}^{( 5 )} ) | - \rangle |$, where $\vec{n} = \vec{0} , N \vec{e}_K$ and $| + \rangle$ and $| - \rangle$ are the eigenkets of $\hat{\sigma}_z$ with eigenvalues $1$ and $-1$, respectively.

If the estimated error $\epsilon_{\rm est}$ is greater than the tolerance $\epsilon_{\rm TOL}$, we change the time step $\delta t$ to the new time step $\delta t_{\rm new}$ defined as \cite{peter2002scientific}
\begin{eqnarray}
\label{eq:NewStepSize}
\delta t_{\rm new} = \left( \frac{0.99 \epsilon_{\rm TOL}}{\epsilon_{\rm est}} \right)^{0.2} \delta t 
\end{eqnarray}
and repeat the one-step evolution with the updated time step. On the other hand, if the estimated error is less than the tolerance, we compute the ADOs using  fifth-order Runge--Kutta. The time step of the next step is determined by Eq.~\eqref{eq:NewStepSize}. 

The parameter values used for the present simulations are summarized in Table~\ref{table:NK}.

\section{TCL-Redfield and Lindblad master equations}
\label{sec:Redfield-master}

\subsection{TCL-Redfield equation}
\label{sec:TCLmaster}
The TCL-Redfield equation is the reduced equation of motion in the case of non-Markovian noise whose damping kernels are expressed in a time-convolution-less form. \cite{shibata1977generalized,chaturvedi1979time,ban2010reduced}
For a spin--boson system,  the equation is expressed as
\begin{equation}
\begin{split}
\frac{\partial}{\partial t} \hat{\rho}_{\rm A} ( t )
&= - \frac{i}{\hbar} \hat{H}_{\rm A}^\times \hat{\rho}_{\rm A} ( t )
- \frac{F_1 ( t )}{\hbar^2} \hat{\sigma}_x^\times \hat{\sigma}_x^\times \hat{\rho}_{\rm A} ( t )
\\
&
\quad - \frac{F_2 ( t )}{\hbar^2} \hat{\sigma}_x^\times \hat{\sigma}_y^\times \hat{\rho}_{\rm A} ( t )
- \frac{i}{\hbar^2} D ( t ) \hat{\sigma}_x^\times \hat{\sigma}_y^\circ \hat{\rho}_{\rm A} ( t ) ,
\end{split}
\end{equation}
where the Redfield tensor elements are defined as
\begin{equation}
\begin{split}
F_1 ( t ) 
&= \frac{A^2 \hbar \gamma^2}{2} \cot \left( \frac{\beta \hbar \gamma}{2} \right)
f_1 ( \gamma ; t ) \\
&
- \sum_{j = 1}^\infty \frac{A^2 \gamma^2}{\beta} 
\frac{2 \nu_j^{\rm (M)}}{\gamma^2 - [ \nu_j^{\rm (M)} ]^2} f_1 ( \nu_j^{\rm (M)} ; t ) ,
\end{split}
\end{equation}
\begin{equation}
\begin{split}
F_2 ( t )
&= \frac{A^2 \hbar \gamma^2}{2} \cot \left( \frac{\beta \hbar \gamma}{2} \right) f_2 ( \gamma ; t ) \\
&
- \sum_{j = 1}^\infty \frac{A^2 \gamma^2}{\beta} \frac{2 \nu_j^{\rm (M)}}{\gamma^2 - [ \nu_j^{\rm (M)} ]^2}
f_2 ( \nu_j^{\rm (M)} ; t ) ,
\end{split}
\end{equation}
and
\begin{eqnarray}
D ( t ) = \frac{A^2 \hbar \gamma^2}{2} f_2 ( \gamma ; t ) ,
\end{eqnarray}
with the functions
\begin{align}
f_1 ( \nu ; t )
= \frac{1}{\omega_0^2 + \nu^2} \left\{ \nu ( 1 - e^{-\nu t} \cos \omega_0 t ) 
+ \omega_0 e^{- \nu t} \sin \omega_0 t \right\}  
\end{align}
and
\begin{align}
f_2 ( \nu ; t ) = \frac{1}{\omega_0^2 + \nu^2}
\left\{ \omega_0 ( 1 - e^{- \nu t} \cos \omega_0 t ) - \nu e^{- \nu t} \sin \omega_0 t  \right\} .
\end{align}
Here, $\nu_j^{\rm (M)} = 2 \pi j / \beta \hbar$ is the $j$th Matsubara frequency and $\omega_0 \equiv 2 B_z / \hbar$ is the characteristic frequency of the subsystem.

\subsection{Lindblad master equation}
\label{sec:Lindblad}

The Lindblad master equation was derived to have Markovian dynamics and to satisfy the detailed balance condition. For a spin--boson system, it is expressed in terms of the creation and annihilation operators $\hat{\sigma}^{+}$ and $\hat{\sigma}^{-}$ as follows: \cite{geva1992quantum}
\begin{equation}
\label{eq:Lindblad}
\begin{split}
\frac{\partial}{\partial t} \hat{\rho}_{\rm A} ( t )
&= -\frac{i}{\hbar} [ \hat{H}_{\rm A} , \hat{\rho}_{\rm A} ( t ) ] \\
&
+ \eta_+ \left[ \hat{\sigma}^+ \hat{\rho}_{\rm A} ( t ) \hat{\sigma}^-
- \frac{1}{2} \left\{ \hat{\sigma}^- \hat{\sigma}^+ , \hat{\rho}_{\rm A} ( t ) \right\} \right] \\
&
+ \eta_- \left[ \hat{\sigma}^- \hat{\rho}_{\rm A} ( t ) \hat{\sigma}^+
- \frac{1}{2} \left\{ \hat{\sigma}^+ \hat{\sigma}^- , \hat{\rho}_{\rm A} ( t ) \right\} \right] .
\end{split}
\end{equation}
Here, $\eta_+$ and $\eta_-$ are chosen to satisfy the detailed balance condition, i.e., $\eta_- / \eta_+ = e^{\beta \hbar \omega_0}$, \cite{geva1992quantum,deffner2019quantum} and,  
for comparison with the SB Hamiltonian system described by Eq.~\eqref{eq:Htotal} for fixed $k$ with the RWA, we define them as
\begin{eqnarray}
\eta_+ = \frac{2 \pi A^2}{\pi} J ( \omega_0 ) n ( \beta ; \omega_0 ) 
\end{eqnarray}
and
\begin{eqnarray}
\eta_- = \frac{2 \pi A^2}{\pi} J ( \omega_0 ) [ n ( \beta ; \omega_0 ) + 1 ] ,
\end{eqnarray}
where $n ( \beta ; \omega ) \equiv 1 / [ \exp({\beta \hbar \omega}) - 1 ]$.

\section{Anomalous behavior of Ohmic spin--boson system}
\label{sec:OhmicLimit}

It is crucial to emphasize that although the Ohmic SDF has been assumed for the study of Markovian dynamics in the spin--boson systems, this model exhibits anomalous dynamical behavior (an infrared anomaly) at finite temperatures.\cite{SpinBosonLeggett,Selbey1984,Cao2019}  We illustrate this point by applying the non-Markovian tests for the spin--boson system (i)--(iii) in Sec.~III of Ref. \onlinecite{KT24JCP3} with a large $\gamma$ to ensure that the Drude SDF [Eq.~\eqref{eq:SDF}] reduces to
\begin{eqnarray}
J ( \omega ) = \frac{\hbar \omega}{\pi}.
\end{eqnarray}
The results are then compared with those obtained using the Lindblad equation (see Appendix~\ref{sec:Lindblad}) instead of the TCL-Redfield equation for comparison. For both calculations, we chose the same condition as in Sec.~\ref{sec:NumericalExamine} besides $\gamma$: we set $\gamma = 100$ for the HEOM calculation. Note that the description of the Lindblad equation has also been investigated for a Brownian oscillator system by comparing its numerical results with exact analytical  solutions.\cite{KT24JCP3}

\begin{table}
\caption{\label{table:EqG} 
Expectation values $\langle \hat{\sigma}_z \rangle = \tr_{\rm A} \{ \hat{\sigma}_z \hat{\rho}_{\rm A}^{\rm eq} \}$ in the equilibrium state computed using the HEOM-MB and the Lindblad equation for $A=1$ in the cases of intermediate
($\beta = 1.0$)  and low ($\beta = 10.0$) temperature.}
\begin{ruledtabular}
\begin{tabular}{cccc}
\hspace{0.3cm} $\beta$ \hspace{0.3cm} & HEOM &  Lindblad
\\
\hline 
$1.0$ & $3.31 \times 10^{-2}$  & $0.762$
\\
$0.1$ & $2.31 \times 10^{-2}$  & $9.97 \times 10^{-2}$
\end{tabular}
\end{ruledtabular}
\end{table}

In Table~\ref{table:EqG}, we list the equilibrium expectation values of $\hat{\sigma}_z$ for  intermediate ($\beta = 1.0$) and low $\beta = 10.0$ temperatures. For the HEOM calculations, we set $N = 8$ and $K = 14$ for $\beta = 1$ and  $N = 9$ and $K = 6$ for $\beta = 0.1$.  The Lindblad master equation is integrated using the fourth-order Runge--Kutta method with  time step $\delta t = 0.001$.

The Lindblad results obtained assuming a factorized and perturbative description of the SB interaction show the Boltzmann distribution of the subsystem itself as the thermal equilibrium state.  By contrast, the HEOM results show significantly smaller excitation populations, even at higher temperatures.  This difference is not due to the RWA assumed in the Lindblad equation, but to the intrinsic anomaly of the coherent elements resulting from the Ohmic SDF.

\begin{figure}
\includegraphics[scale=0.4]{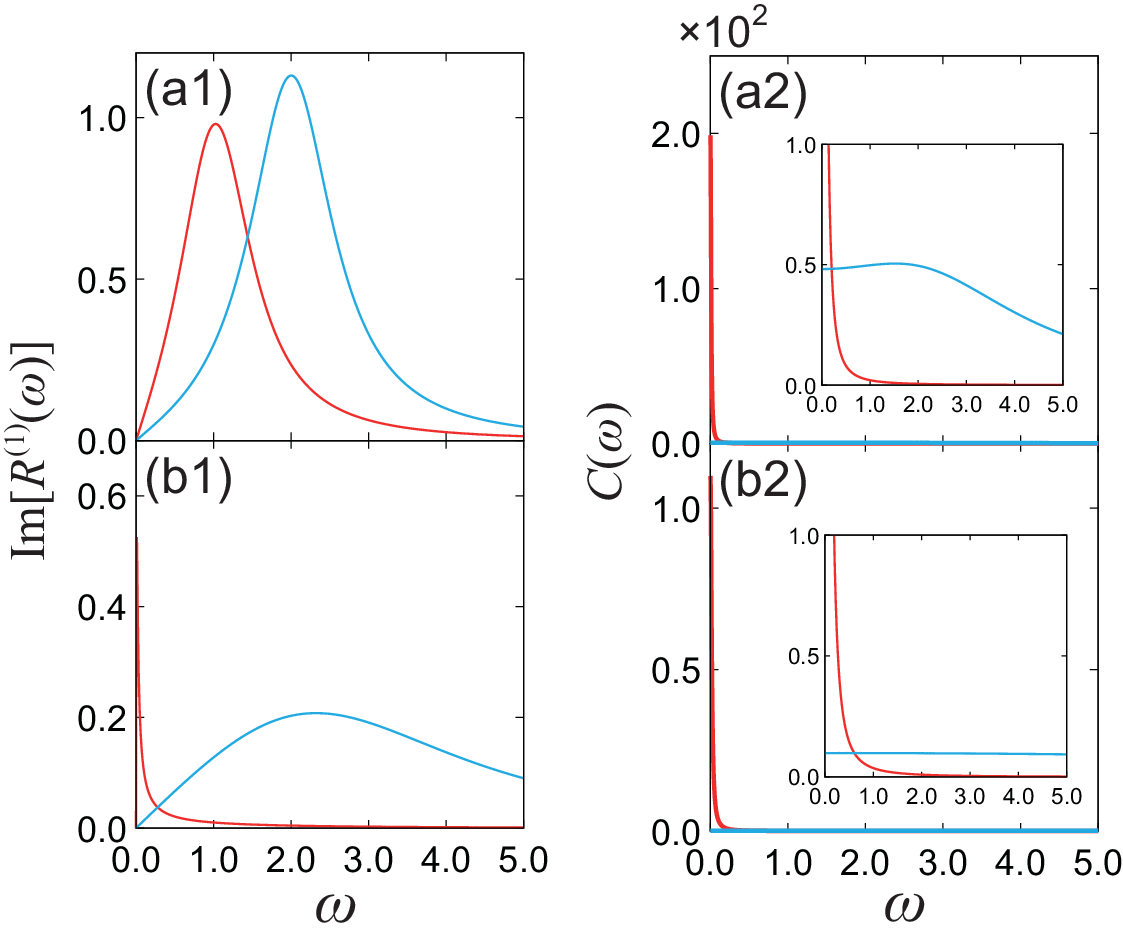}
\caption{\label{fig:G100}Fourier elements of (1) the linear response function $R^{(1)} [ \omega ]$ and (2) the symmetric correlation function  $C [\omega]$ for (a1) weak  SB coupling ($A = 0.5$) and (b1) intermediate SB coupling ($A = 1.0$) at intermediate temperature ($\beta =1.0$) and for
(a2)  intermediate temperature  ($\beta = 1.0$) and (b2)  low temperature ($\beta = 10.0$) at $A=0.1$. In each plot, the red and green curves represent the HEOM and Lindblad results, respectively. In (a2) and (b2), the lower-frequency region of the spectrum  is enlarged and shown as an inset. }
\end{figure}

In Fig.~\ref{fig:G100}, we present (1) the Fourier element of the linear response function $R^{(1)} ( t ) = i  \tr_{\rm tot} \{ [ \hat{\sigma}_x ( t ) , \hat{\sigma}_x ] \hat{\rho}_{\rm tot}^{\rm eq} \}/\hbar$ at intermediate temperature ($\beta =1.0$) for (a1)  weak SB coupling [$A ( t ) = 0.5$]   and (b1)  intermediate  SB coupling [$A ( t ) = 1.0$] and (2) the Fourier element of the symmetric correlation function $C ( t ) = \tr_{\rm tot} \{ \{ \hat{\sigma}_x ( t ), \hat{\sigma}_x \} \hat{\rho}_{\rm tot}^{\rm eq} \} / 2$ for $A=1$ at (a2) intermediate temperature ($\beta = 1.0$) and (b2) low temperature ($\beta = 10.0$). 
The red and green curves represent the results from the HEOM and the Lindblad equation, respectively. 
To perform the HEOM calculations, we set $N = 7$ and $K = 14$ in (a1), $N = 9$ and $K = 8$ in (b1), $N = 9$ and $K = 8$ in (a2), and $N = 9$ and $K = 6$ in (b2).

The peak positions of the HEOM results in the nearly Ohmic case are shifted toward zero frequency compared with Fig.~\ref{fig:LRFCF} for the non-Markovian case. 
This is because the effect of the fluctuation on the coherent elements $\langle \hat{\sigma}_x \rangle = \tr_{\rm A} \{ \hat{\sigma}_x \hat{\rho}_{\rm A} ( t ) \}$, given by the first term on the right-hand side of Eqs.~\eqref{eq:Theta0} and~\eqref{eq:Thetal}, approaches infinity in the Ohmic limit $\gamma \rightarrow \infty$, as was shown in the case of the low-temperature Fokker--Planck equation.\cite{IT19JCTC}
Since the Lindblad master equation approach completely ignores these effects, the spectra evaluated using this approach differ significantly from the actual ones.  As shown in Table~\ref{table:EqG}, the difference in the coherent elements further appears as a difference in the population distribution. The reason why this effect is not pronounced in actual experimental results is that the noise in a real system is not Markovian; special care must be taken when using the Lindblad master equation to analyze experimental results.

\section{Evaluation of the entropy production}
\label{sec:ComputePlanck}

From Eq.~\eqref{eq:Planck-Gibbs}, we can evaluate the quasi-static Planck potential $\Xi_{\rm tot}^{\rm qst} ( t )$ from the Gibbs energy. Then, by evaluating $\tilde{W}_{\rm tot}^{int}$, we can obtain the entropy production from Eq. \eqref{eq:loss}.
Here, we consider the time-dependent SB coupling in the isothermal case to evaluate $\tilde{W}_{\rm tot}^{int}$ and $\Sigma_{\rm tot}$:
\begin{equation}
\label{eq:ADetachment}
A ( t ) = 
\begin{cases}
A_0 & ( t < 0 ),
\\
A_0 ( 1 - t / \tau ) & ( 0 \leq t < \tau ),
\\
0 & ( \tau \leq t ),
\end{cases}
\end{equation}
where $\tau$ is the time duration of the bath detachment process and $A_0$ is the initial SB coupling strength. We assume that the Hamiltonian of the system is time-independent. From the Kelvin--Planck statement, when the process is quasi-static, the intensive work performed in the above process is equal to the change in the quasi-static Gibbs energy $\Delta G^{\rm qst}$. In this section, we omit the subscript ``tot'' of the Gibbs energy.
In the limit of weak SB coupling  $A ( t ) \rightarrow 0$, the effect of  bathentanglement becomes negligible, and the reduced density operator of the subsystem is described by the Boltzmann distribution  $\hat{\rho}_{\rm A}^{\rm eq} = e^{- \beta \hat{H}_{\rm A}} / Z_{\rm A}$, where $Z_{\rm A} = \tr_{\rm A} \{ e^{- \beta \hat{H}_{\rm A}} \}$ is the partition function of the subsystem; the Gibbs energy in the weak-SB-coupling limit is expressed as $G^{\rm qst}_{A \rightarrow 0} = - \beta^{-1} \ln Z_{\rm A}$.

Using  $\Delta G^{\rm qst}$ and $G^{\rm qst}_{A \rightarrow 0}$, we can calculate $G^{\rm qst}_{A_0}$  as
\begin{eqnarray}
G^{\rm qst}_{A_0} = G^{\rm qst}_{A \rightarrow 0} - \Delta G^{\rm qst} .
\end{eqnarray}

To perform numerical simulation, we set $\tau = 1.0 \times 10^4$ in Eq.~\eqref{eq:ADetachment}.  The results are summarized and displayed in table \ref{table:DLWint}.

\bibliography{references,tanimura_publist}

\end{document}